\documentclass{mn2e}
\include{epsf}

\def \farcs{\hbox{$.\!\!^{\prime\prime}$}}
\def \farcm{\hbox{$.\!\!^{\prime}$}}
\def \msun{\hbox{M$_\odot$}}

\title[Weak lensing masses of clusters]{Comparison of weak lensing
masses and X-ray properties of galaxy clusters\thanks{Based on observations 
from the Canada-France-Hawaii Telescope, which is operated by the National
Research Council of Canada, le Centre National de la Recherche
Scientifique and the University of Hawaii.}}

\author[H. Hoekstra]{Henk Hoekstra\\
Department of Physics and Astronomy, University of
Victoria, Victoria, BC, V8P 5C2, Canada}

\begin{document}

\date{Accepted. Received; in original form}

\maketitle

\begin{abstract}

We present measurements of the masses of 20 X-ray luminous clusters of
galaxies at intermediate redshifts, determined from a weak lensing
analysis of deep archival $R$-band data obtained using the
Canada-France-Hawaii-Telescope. Compared to previous work, our
analysis accounts for a number of effects that are typically ignored,
but can lead to small biases, or incorrect error estimates. We derive
masses that are essentially model independent and find that they agree
well with measurements of the velocity dispersion of cluster galaxies
and with the results of X-ray studies. Assuming a power law between
the lensing mass and the X-ray temperature, $M_{2500}\propto
T^\alpha$, we find a best fit slope of
$\alpha=1.34^{+0.30}_{-0.28}$. This slope agrees with self-similar
cluster models and studies based on X-ray data alone. For a cluster
with a temperature of $kT=5$ keV we obtain a mass
$M_{2500}=(1.4\pm0.2)\times 10^{14}h^{-1}\msun$ in fair agreement with
recent Chandra and XMM studies.

\end{abstract}


\section{Introduction}

Clusters of galaxies have been the focus of intense study for many
decades for a variety of reasons. They are the largest gravitationally
bound objects in the universe and provide a large reservoir of
baryons, mainly in the form of the hot intracluster (ICM) gas.  The
study of galaxy clusters has been transformed recently with the advent
of powerful X-ray telescopes such as Chandra and XMM, which enable us
to determine the properties of the gas in unprecedented detail and
accuracy. Such studies will hopefully lead to a better understanding
of the complex physics of the ICM and its interaction with the various
constituents of the cluster (e.g., dark matter, galaxies). The
dynamical state of the ICM might also reveal information about the
recent merger history, thus providing an interesting way to test the
concept of hierarchical cluster formation, as predicted in the cold
dark matter (CDM) paradigm.

\begin{table*}
\begin{center}
\caption{Summary of the observational data for the cluster sample\label{tabsample}}
\begin{tabular}{llllrcrc}
\hline
\hline
name      & RA  & DEC & $z$  & $t_{\rm exp}$(B) & seeing(B) & $t_{\rm exp}$(R) & seeing(R)\\
          &     &     &      & [s] & [arcsec] & [s] & [arcsec] \\
\hline
A2390      & $21^h53^m36.8^s$ & $+17^\circ41'44''$ & 0.2280 & 8100  & 0.90 & 9600  & 0.66 \\
MS 0016+16 & $00^h18^m33.5^s$ & $+16^\circ26'16''$ & 0.5465 & 6180  & 0.87 & 9423  & 0.68 \\
MS 0906+11 & $09^h09^m12.6^s$ & $+10^\circ58'28''$ & 0.1704 & 6000  & 0.93 & 13800 & 0.84 \\
MS 1224+20 & $12^h27^m13.5^s$ & $+19^\circ50'56''$ & 0.3255 & 8640  & 1.01 & 11400 & 0.86 \\
MS 1231+15 & $12^h33^m55.4^s$ & $+15^\circ25'58''$ & 0.2353 & 3900  & 0.90 & 9000  & 0.71 \\
MS 1358+62 & $13^h59^m50.6^s$ & $+62^\circ31'05''$ & 0.3290 & 11880 & 1.04 & 6300  & 0.82 \\
MS 1455+22 & $14^h57^m15.1^s$ & $+22^\circ20'35''$ & 0.2568 & 4800  & 1.03 & 9000  & 0.67 \\
MS 1512+36 & $15^h14^m22.5^s$ & $+36^\circ36'21''$ & 0.3727 & 12540 & 0.90 & 9900  & 0.71 \\
MS 1621+26 & $16^h23^m35.5^s$ & $+26^\circ34'14''$ & 0.4275 & 15300 & 0.99 & 11520 & 0.72 \\
\hline
A68        & $00^h37^m06.9^s$ & $+09^\circ09'24''$ & 0.255  & 8100  & 1.05 & 7200  & 0.67 \\
A209       & $01^h31^m52.5^s$ & $-13^\circ36'40''$ & 0.206  & 7200  & 0.98 & 5400  & 0.70 \\
A267       & $01^h52^m42.0^s$ & $+01^\circ00'26''$ & 0.230  & 3000  & 0.93 & 4800  & 0.72 \\
A383       & $02^h48^m03.4^s$ & $-03^\circ31'44''$ & 0.187  & 7200  & 0.93 & 4800  & 0.90 \\
A963       & $10^h17^m03.8^s$ & $-39^\circ02'51''$ & 0.206  & 7200  & 0.88 & 4800  & 0.77 \\
A1689      & $13^h11^m30.0^s$ & $-01^\circ20'30''$ & 0.1832 & 3600  & 0.88 & 3000  & 0.81 \\
A1763      & $13^h35^m20.1^s$ & $+41^\circ00'04''$ & 0.223  & 3600  & 0.94 & 6000  & 0.85 \\
A2218      & $16^h35^m48.8^s$ & $+66^\circ12'51''$ & 0.1756 & 3378  & 1.06 & 3300  & 0.84 \\
A2219      & $16^h40^m19.9^s$ & $+46^\circ42'41''$ & 0.2256 & 5400  & 0.91 & 6300  & 0.78 \\
\hline
A370       & $02^h39^m52.7^s$ & $-01^\circ34'18''$ & 0.375  & 10408 & 0.92 & 10800 & 0.77 \\
CL0024+16  & $00^h26^m35.6^s$ & $+17^\circ09'44''$ & 0.390  & 12960 & 0.98 & 9000  & 0.64 \\
\hline
\hline
\end{tabular}

\bigskip

\begin{minipage}{0.75\linewidth}

{\footnotesize Column 1: cluster name; Column 2,3: right ascension and
declination (J2000.0) of the brightest cluster galaxy. Note that in
the case of two dominant central galaxies this might differ from
previous positions; Column 4: cluster redshift; Column 5,7: exposure
times in the $B$ and $R$-band resp.; Column 6,8: seeing in $B$ and $R$
resp. The top 9 clusters have been studied as part of the CNOC cluster
survey. The next 9 systems are a subset of the systems studied by
Bardeau et al. (2005) (although we note that A2390 is also part of that
study). The last two clusters are well known clusters because of their
spectacular strong lensing.}

\end{minipage}

\end{center}
\end{table*}

As the most massive objects in the universe, galaxy clusters are
readily found out to large redshifts in optical, X-ray or millimeter
(Sunyaev-Zel'dovich; SZ) surveys. This makes them excellent
cosmological probes, because the number of clusters as a function of
mass is very sensitive to a range of cosmological parameters, such as
the matter density, the normalisation of the matter power spectrum and
the equation of state of the dark energy (e.g., Evrard 1989; Eke et
al. 1998; Henry 2000; Haiman et al. 2001; Levine et al. 2002; Allen et
al. 2004). These constraints are complimentary to measurements of
large scale structure and the cosmic microwave background.

Given the great interest in the use of galaxy clusters as a
cosmological probe, and the advent of efficient SZ telescopes in the
coming years it is important to derive accurate masses for these
systems. To estimate masses from the motion of cluster galaxies, one
needs to make assumptions about the orbital structure and the geometry
of the cluster in addition to the assumption of equilibrium.
Similarly, X-ray and SZ measurements of the total mass assume the
cluster is in hydrostatic equilibrium. These assumptions are not
always valid, thus complicating systematic studies of cluster
properties. Finally, it is becoming increasingly clear that
non-gravitational physics complicates matters further.

Fortunately, there exists a direct way to determine the cluster
mass. The gradient in the gravitational potential of the cluster
causes differential deflection of light rays coming from distant
galaxies.  This causes small, systematic distortions in the shapes of
these faint sources, an effect known as weak gravitational lensing.
The amplitude of the signal provides a direct measurement of the
projected mass {\it along the line of sight} in a given aperture, which in
turn can be compared directly to numerical simulations, a crucial step
if one wants to use clusters for cosmology.  However, to compare the
weak lensing results to other mass indicators one typically has
to make assumptions regarding the cluster geometry.

Weak lensing is now a well established technique to study the
distribution of (dark) matter in the universe and the applications are
numerous (e.g., see Hoekstra et al., 2002a; Schneider 2005). In particular,
recent progress in the measurement of the lensing signal caused by
large scale structure (cosmic shear) has demonstrated the reliability
with which the tiny distortions in the images of faint galaxies can be
measured (e.g., Hoekstra et al. 2002b; Van Waerbeke et al. 2005;
Hoekstra et al. 2006).

Although the first succesful detection of weak lensing was made
studying a galaxy cluster (Tyson, Wenk \& Valdes 1990), the sample of
clusters with accurate mass determinations is relatively small. Early
results are based on inhomogeneous data sets which cover only the
central regions (e.g., Fahlman et al. 1994; Hoekstra et al. 1998;
Squires et al. 1996a, 1996b). More recently, Dahle et al. (2002) 
and Cypriano et al. (2005) published results on larger samples.
This situation has limited detailed comparisons of masses inferred
from X-ray observations and those of lensing, although some attempts
have been made (e.g., Allen 1998).

In recent years it has become possible to study the mass distribution
of clusters in much more detail, thanks to the advent of wide field
imagers. These instruments enable us to measure the lensing signal out
to much larger radii, improving the reliability of the results.
In addition, thanks to studies of photometric redshifts for the faint
galaxies used in weak lensing analysis, we can now better relate the
observed lensing signal to an estimate of the cluster mass.

In this paper we present measurements of the weak lensing masses for a
sample of 20 X-ray luminous clusters of galaxies observed with the
CFH12k camera on the Canada-France-Hawaii Telescope (CFHT). The data
used here consist of archival $B$ and $R$ band images, where the
latter data are used in the weak lensing analysis. 

The structure of the paper is as follows. In \S2 we present the data
and discuss the image processing. \S3 deals with the weak lensing
shape measurements and interpretation of the results. The mass
measurements are presented in \S4. In this section we also compare our
results to those obtained from other techniques. Throughout the paper
we assume a cosmology with $\Omega_m=0.3$, $\Omega_\Lambda=0.7$ and
$H_0=100 h$~km/s/Mpc.

\section{Data}

We searched the CFHT archive at the Canadian Astronomical Data Centre
(CADC) to find galaxy clusters that have deep exposures in both the
$B$ and the $R$-band obtained using the CFH12k camera. This camera
consists of an array of 6 by 2 CCDs, each 2048 by 4096 pixels. The
pixel scale is $0\farcs206$, which ensures good sampling for the
subarcsecond imaging data used here. The resulting field of view is
about 42 by 28 arcminutes, significantly larger than the $\sim8 \times
8$ arcminutes typically used in older lensing studies.

This search yielded a sample of 20 X-ray luminous clusters for which
useful data could be retrieved. Nine of the clusters were observed as
part of a follow-up project (PI: P. Fischer) to image the clusters
studied in the Canadian Network for Observational Cosmology Cluster
Redshift Survey (CNOC1; e.g., Yee et al., 1996; Carlberg et al.,
1996). The latter survey is a multi-object spectroscopy survey of
galaxies in moderately rich to rich clusters at intermediate redshifts
$(0.17<z<0.55)$ selected from the Einstein Medium Sensitivity Survey
(EMSS; Gioia et al., 1990). Another 9 clusters in our sample were
taken for another project aiming to determine weak lensing masses
(PIs: Kneib, Czoske). See Bardeau et al (2005) and Smith et al. (2005)
for a discussion of the $z=0.2$ cluster sample, which is selected from
the X-ray Brightest Abell Clusters Survey (XBACS; Ebeling et al.,
1996). Abell 1835 is not studied here because of the lack of deep
$B$-band imaging data. Also note that Abell 2390 is part of both
samples. Finally, two additional clusters at somewhat higher redshifts
are included. These clusters have been studied because of their
exeptional strong lensing properties, and therefore may be considered
``lensing-selected''.

Table~\ref{tabsample} lists the total integration times in the $B$ and
$R$ band for the selected exposures, as well as the seeing measured
from the stacked images. Because the weak lensing analysis requires
good image quality, we omitted poor seeing images when combining the
data. We only use the $R$ band data for the lensing analysis. The $B$
band data are used to select cluster galaxies, and consequently we do
not require excellent seeing for the $B$ band data. The image quality
of the $R$ band data is excellent for most of the images.

\subsection{Image processing}

Detrended data (de-biased and flatfielded) are provided to the
community through CADC. The detrending is done using the Elixir
pipeline developed at CFHT. The pipeline also provides photometric
zeropoints and in most cases a reasonable first order astrometric
solution.

For the weak lensing analysis we use a stack of the individual
exposures. This requires that the astrometry of the images be very
precise. If this were not the case, the misalignment of images would
introduce a spurious signal. Related to this is the need to remove
the distortion of the camera, which places requirements on the overall
astrometry, which are readily met using the procedure outlined below.

In principle one could use the USNO-A2 catalog to refine the
astrometry, but the number density of sources is often too low to
warrant stable results. Instead, we retrieve red images from the
second generation Digital Sky Survey (POSS II) for each cluster. These
observations have small geometric distortions. This can be taken care
of by calibrating the astrometry of the POSS II images using the
USNO-A2 catalog. SExtractor (Bertin \& Arnouts 1996) is used to
generate a catalog of sources with accurate astrometry, with a number
density significantly higher than the USNO-A2 catalog. In addition,
the POSS II images have been taken more recently, thus reducing the
effects of proper motions of the stars.

This new astrometric catalog is matched to each of the CFH12k
exposures. We combine the matched catalogs for each exposure into a
master catalog, which contains the average positions of the matched
objects. This master catalog is used to derive the final second order
astrometric solution for each chip. This procedure ensures that in the
overlapping areas the objects are accurately matched to the same
position.

The images with the improved astrometry are stacked using the SWarp
routine into a large mosaic image. As discussed below, for certain
regions, data from different chips contribute to the image. If the PSF
properties are discontinuous between chips, this may result in
complicated behaviour of the PSF. Fortunately, for most of the data,
the offsets between exposures are small. In addition, the CFH12k PSF
shows no evidence of ``jumps'', because the chips are very well
aligned (e.g., see Hoekstra et al. 2002b).

As mentioned earlier, the Elixir processed images contain photometric
calibrations based on observations of standard stars during the
observing run. These zero-points are only valid under photometric
conditions, a condition that needs to be checked for each exposure.
We therefore examine the magnitudes of a large number of objects in
the images, which enables us to examine the variation in the
photometric zero-point. We found that most data were taken under
photometric conditions. For a few clusters only part of the data were
photometric and we use these images to scale the non-photometric data.

As additional checks we compared galaxy counts (with the cluster
region excised) to those provided by P. Hsieh (private communication),
and the expected distribution of $B-R$ colors of stars. In both cases
we found good agreement between our data and the reference data.
Where available we also compared our magnitudes to those observed by
the Sloan Digital Sky Survey, and found excellent agreement.  As a
final check, we note that colors of the cluster red-sequence are in
good agreement with the expected values (the scatter around the mean
is 0.03 in $B-R$).

\section{Weak lensing analysis}

In this section we discuss the details of the weak lensing analysis
and how we interpret the resulting lensing signal. In \S3.1 we discuss
how we measure the shapes of galaxies used in the weak lensing
analysis. A useful way of quantifying the lensing signal is presented
in \S3.2. There are various ways to estimate the cluster mass from the
observed lensing signal. For instance, one can adopt a parameterized
model for the density profile and fit this to the observations. We
consider two such models. The singular isothermal sphere is discussed
in \S3.3 and the cold dark matter NFW profile (Navarro, Frenk \& White
1997; NFW hereafter) is presented in \S3.4. Another approach is to
determine the projected mass within an aperture of a given radius,
with a minimal dependence on the density profile at large radii. This
approach is discussed in \S3.5.

Various effects that complicate a simple interpretation of the lensing
signal are reviewed in \S3.6. In section \S3.7 we discuss the
contamination of the lensing signal by cluster members, which are
included in the catalog of source galaxies. The conversion of the
lensing signal into a mass requires knowledge of the redshift
distribution of the source galaxies, which is addressed in
\S3.8. Finally, lensing is sensitive to all matter along the line of
sight, which needs to accounted for. This is reviewed in \S3.9.

\subsection{Shape measurements}

The various steps in the weak lensing analysis have been described in
detail in Hoekstra et al. (1998, 2000). Our method is based on the
original procedure developed by Kaiser, Squires \& Broadhurst (1995)
and Luppino \& Kaiser (1997) with modification presented by Hoekstra
et al. (1998). The key step in the lensing analysis is to accurately
measure the shapes of the faint background galaxies, correcting for
the various observational distortions, such as seeing and PSF
anisotropy.

For the weak lensing analysis we only use the stacked $R$-band
images. As can be seen from Table~\ref{tabsample} the $R$ exposures
have better image quality than the $B$-band data. Good image quality
is crucial for an accurate measurement of the mass using weak lensing.

\begin{figure}
\begin{center}
\leavevmode
\hbox{%
\epsfxsize=8.5cm
\epsffile{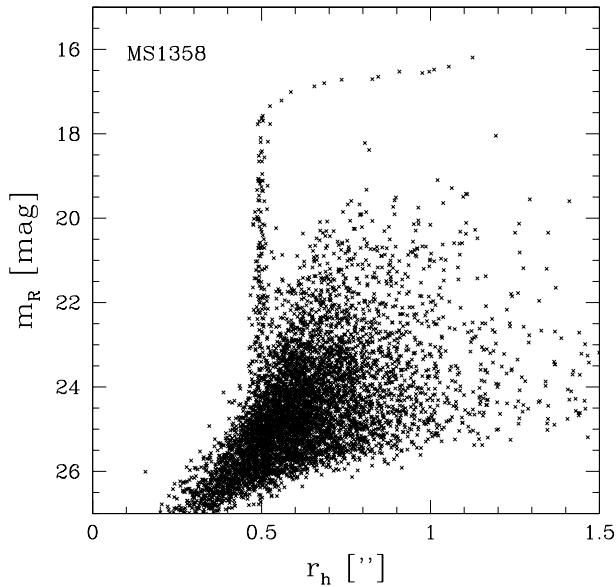}}
\caption{Plot of the apparent magnitude as a function
of the measured half-light radius for the data for MS~1358. The
clear vertical locus of stars at a half-light radius of $\sim 0\farcs5$ 
allows for a clean selection of stars used to characterize the PSF.}
\label{sizemag}
\end{center}
\end{figure}

Although the PSF anisotropy varies coherently over the total field of
view, we follow van Waerbeke et al. (2005) and split the coadded
images into subsets which correspond approximately to the original
chips. This enables us to better characterize the spatial variation of
the PSF anisotropy. The various images used to obtain the deep coadded
images have been taken with (typically) small offsets, which in
principle complicates the analysis. However, discontinuities in the
PSF anisotropy between chips are small and can be neglected in our
analysis. This is supported by the results from the VIRMOS-DESCART
survey (van Waerbeke et al. 2005). These cosmic shear measurements are
very sensitive to residual systematics, but do not show a significant
systematic signal due to the adopted approach. We conclude by noting
that the weak lensing analysis of galaxy clusters is less sensitive to
residual systematics compared to cosmic shear studies.

The first step in the lensing analysis is the identification of the
faint galaxies in the images.  For that we use the hierarchical peak
finding algorithm from Kaiser et al. (1995) to find objects with a
significance $>5\sigma$ over the local sky. The peak finder gives fair
estimates for the object size, and we reject all objects smaller than
the size of the PSF.

The objects in this cleaned catalog are analysed, which yields
estimates for the size, apparent magnitude and shape parameters
(polarisation and polarisabilities) and their measurement errors. The
objects in this catalog are inspected by eye, in order to remove
spurious detections. These objects have to be removed because their
shape measurements are affected by cosmetic defects (such as bleeding
stars, diffraction spikes from bright stars) or because the objects
are likely to be part of a resolved galaxy (e.g., an HII region).

To measure the small, lensing induced distortions it is important to
accurately correct the shapes for PSF anisotropy, as well as for the
dilluting effect of seeing. To characterize the spatial variation of
the PSF we select a sample of moderately bright stars from our images.
The stars are selected based on their location in a plot of the
apparent magnitude as a function of half-light radius, such as the one
presented in Figure~\ref{sizemag}. The stars are defined by the
vertical locus at $r_h=0\farcs5$, and allow for a clean sample of
stars. We fit a second order polynomial to the shape parameters of the
selected stars for each chip, which provides us with a estimate of the
PSF anisotropy at every position. These results are used to correct
the shapes of the galaxies for PSF anisotropy.

Seeing circularizes the images, thus lowering the raw lensing signal.
To correct for the seeing, we need to rescale the polarisations to
their ``pre-seeing'' value. This scale factor is the ``pre-seeing''
shear polarisability $P^\gamma$ (Luppino \& Kaiser 1997; Hoekstra et
al. 1998). The measurements of $P^\gamma$ for each individual galaxy
are noisy, and we therefore bin the measurements as a function of
galaxy size, and use the ensemble average value as a function of size
to correct for the effect of seeing. This approach has been tested on
simulated images in great detail (e.g., Hoekstra et al. 1998; Heymans
et al. 2005).  These results suggest that we can recover the shear
with an accuracy of $\sim 2\%$ (Heymans et al. 2005).

Having corrected the shapes of the galaxies in each ``chip'', we
combine the catalogs into a master catalog, which covers the full
field of view. This catalog is used for the weak lensing analysis.

\subsection{Tangential distortion}

The azimuthally averaged tangential shear $\langle \gamma_t\rangle$ as
a function of radius from the cluster centre is a useful measure of
the lensing signal (e.g., Miralda-Escud{\'e} 1991): 

\begin{equation}
\langle\gamma_t\rangle(r)=\frac{\bar\Sigma(<r) - \bar\Sigma(r)}
{\Sigma_{\rm crit}}=\bar\kappa(<r)-\bar\kappa(r),
\end{equation}

\noindent where $\bar\Sigma(<r)$ is the mean surface density within an
aperture of radius $r$, and $\bar\Sigma(r)$ is the mean surface
density on a circle of radius $r$. The convergence $\kappa$, or
dimensionless surface density, is the ratio of the surface density and
the critical surface density $\Sigma_{\rm crit}$, which is given by

\begin{equation}
\Sigma_{\rm crit}=\frac{c^2}{4\pi G}\frac{D_s}{D_l D_{ls}},
\end{equation}

\noindent where $D_l$ is the angular diameter to the lens. $D_{s}$ and
$D_{ls}$ are the angular diameter distances from the observer to the
source and from the lens to the source, respectively. The parameter
$\beta=D_{ls}/D_s$ is a measure of how the amplitude of the lensing
signal depends on the redshifts of the source galaxies.

\begin{figure}
\begin{center}
\leavevmode
\hbox{%
\epsfxsize=8.5cm
\epsffile{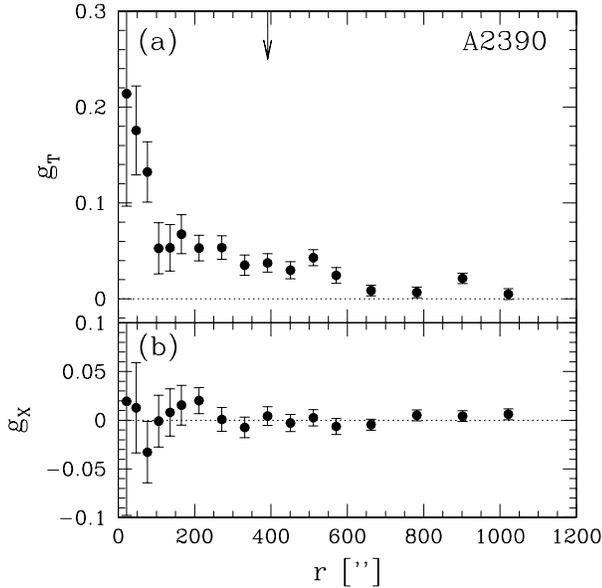}}
\caption{{\it Panel (a):} plot of the tangential distortion as a
function of the distance from the cluster center for Abell 2390 using
galaxies with $21<R<25$. A significant signal is measured out to large
radii. The arrow indicates a radius of $1 h^{-1}$ Mpc at the redshift
of the cluster.  {\it Panel (b):} the signal when the sources are rotated by 45
degrees. If the signal presented in panel~a is caused by lensing, no
signal should be present, as is the case. This test is similar to the
``B''-modes in cosmic shear surveys.}
\label{gtprof}
\end{center}
\end{figure}

Figure~\ref{gtprof}a shows the observed lensing signal as a function
of distance from the cluster center for Abell 2390. The center is taken to
be the position of the brightest cluster galaxy. A significant signal
is measured out to large radii. Panel~b indicates a test for
systematics: if the signal presented in panel~a is caused by lensing,
no signal should be present, as is observed. To minimize the
contamination by cluster members, we have excluded galaxies with
colors similar to the cluster color-magnitude relation (see \S3.7 for
a more detailed discussion).

\subsection{Singular Isothermal Sphere}

A simple model for the cluster mass distribution is the singular
isothermal sphere (SIS) for which the convergence and tangential shear
are given by

\begin{equation}
\kappa=\langle\gamma_t\rangle=\frac{r_E}{2r},
\end{equation}

\noindent where $r_E$ is the Einstein radius. For the clusters in our
sample we fit this model to the observed lensing signal (taking into
account the complicating factors discussed below) from 0.25 to $1
h^{-1}$Mpc, and the resulting values with their statistical errors are
listed in Table~\ref{tabsis}.

Under the assumption of isotropic orbits and spherical symmetry, the
Einstein radius (in radians) is related to the line-of-sight velocity
dispersion through

\begin{equation}
r_E=4\pi\left(\frac{\sigma}{c}\right)^2\beta,
\end{equation}

\noindent which allows for a direct comparison with measurements of
the line-of-sight velocity dispersion of cluster galaxies from
redshift surveys. We compare our lensing results with these dynamical
measurements in \S4.2.

\begin{table*}
\caption{Results for the singular isothermal sphere model \label{tabsis}}
\centering
\begin{tabular}{lccccccccc}
\hline
\hline
(1)  & (2) & (3) & (4) & (5) & (6) & (7) & (8) & (9) & (10)\\
name       & $R_C$ & $\langle\beta\rangle$ & $\langle\beta^2\rangle$ & $D_l$          
& $1 h^{-1}$ Mpc & $r_E$ & $\sigma_{r_E}$ & $\sigma_{\rm LSS}$ & $\sigma_{\rm tot}$ \\
           & [mag] &         &     & [$h^{-1}$ Gpc] & ['']  & ['']  & ['']  & [''] & [''] \\ 
\hline
A2390      & 21-25   & 0.54 & 0.34 & 0.53 & 391 & 19.4 & 2.3 & 1.3 & 2.6 \\
MS 0016+16 & 22-25.5 & 0.24 & 0.11 & 0.92 & 223 &  9.2 & 2.2 & 1.0 & 2.4 \\
MS 0906+11 & 21-25   & 0.62 & 0.43 & 0.42 & 492 & 13.9 & 2.7 & 1.6 & 3.2 \\
MS 1224+20 & 21-25   & 0.40 & 0.21 & 0.68 & 303 &  8.1 & 2.4 & 1.1 & 2.7 \\
MS 1231+15 & 21-25.5 & 0.54 & 0.35 & 0.54 & 382 &  5.0 & 2.3 & 1.5 & 2.8 \\
MS 1358+62 & 21-25   & 0.40 & 0.22 & 0.69 & 301 & 12.6 & 2.2 & 1.1 & 2.5 \\
MS 1455+22 & 21-25.5 & 0.52 & 0.33 & 0.58 & 358 & 13.9 & 2.1 & 1.4 & 2.5 \\
MS 1512+36 & 22-25.5 & 0.40 & 0.22 & 0.74 & 278 &  5.9 & 2.2 & 1.2 & 2.5 \\
MS 1621+26 & 22-25.5 & 0.34 & 0.18 & 0.81 & 255 &  9.7 & 2.3 & 1.1 & 2.5 \\
\hline
A68        & 21-25   & 0.49 & 0.29 & 0.57 & 360 & 15.2 & 2.3 & 1.3 & 2.6 \\
A209       & 21-25   & 0.57 & 0.38 & 0.49 & 423 & 13.3 & 2.3 & 1.4 & 2.7 \\
A267       & 20-25   & 0.53 & 0.33 & 0.53 & 389 & 15.5 & 2.4 & 1.3 & 2.8 \\
A383       & 21-24.5 & 0.58 & 0.38 & 0.45 & 456 &  8.2 & 3.2 & 1.3 & 3.4 \\
A963       & 21-25   & 0.57 & 0.37 & 0.49 & 423 & 11.6 & 2.4 & 1.4 & 2.8 \\
A1689      & 21-24.5 & 0.60 & 0.40 & 0.44 & 464 & 32.2 & 2.7 & 1.3 & 2.9 \\
A1763      & 21-25   & 0.54 & 0.33 & 0.52 & 398 & 17.3 & 2.5 & 1.4 & 2.9 \\
A2218      & 21-24.5 & 0.61 & 0.42 & 0.43 & 480 & 19.1 & 2.9 & 1.3 & 3.2 \\
A2219      & 21-25   & 0.54 & 0.34 & 0.52 & 394 & 17.9 & 2.4 & 1.4 & 2.7 \\
\hline 
A370       & 22-25   & 0.37 & 0.19 & 0.75 & 277 & 19.7 & 2.4 & 1.0 & 2.6 \\
CL0024+16  & 22-25.5 & 0.36 & 0.19 & 0.76 & 270 & 13.6 & 2.4 & 1.2 & 2.7 \\
\hline
\hline
\end{tabular}

\bigskip

\begin{minipage}{0.75\linewidth}

{\footnotesize Column 1: cluster name; Column 2: adopted range in
apparent magnitude for the source galaxies; Column 3: effective value
of $\langle\beta\rangle$ as described in the text. The derived
statistical error is 0.01; Column 4: value for
$\langle\beta^2\rangle$; Column 5: angular diameter distance to the
cluster; Column 6: angular size corresponding to $1h^{-1}$ Mpc; Column
7: value for the Einstein radius $r_E$ obtained from a SIS model fit
to the tangential distortion from 0.25 to 1.5 $h^{-1}$ Mpc; Column 8:
statistical error for the measurement of the Einstein radius; Column
9: uncertainty in $r_E$ caused by uncorrelated large scale structure
along the line of sight; Column 10: total uncertainty in $r_E$,
combining the statistical and LSS errors.}

\end{minipage}
\end{table*}

\subsection{NFW profile}

Collisionless cold dark matter (CDM) provides a good description for
the observed structure in the universe.  Numerical simulations,
indicate that on large scales CDM gives rise to a particular density
profile (e.g., Dubinski \& Carlberg 1991; Navarro, Frenk, \& White
1995; Navarro, Frenk, \& White 1997; Moore et al. 1999). In these
simulations, the NFW profile, given by

\begin{equation}
\rho(r)=\frac{M_{\rm vir}}{4\pi f(c)}\frac{1}{r(r+r_s)^2},
\end{equation}

\noindent appears to be an good description of the radial mass
distribution for halos with a wide range in mass. Here, $M_{\rm vir}$
is the virial mass, which is the mass enclosed within the radius
$r_{\rm vir}$. The virial radius is related to the scale radius
through the concentration $c=r_{\rm vir}/r_s$, and the function
$f(c)=\ln(1+c)-c/(1+c)$.

One can fit the NFW profile to the measurements with $M_{\rm vir}$ and
concentration $c$ (or equivalently $r_s$) as free parameters. However,
numerical simulations have shown that the average concentration
depends on the halo mass and the redshift (Bullock et al. 2001)

\begin{equation}
c=\frac{9}{1+z}\left(\frac{M_{\rm vir}}{8.12\times 10^{12} h
M_\odot}\right)^{-0.14}.
\end{equation}

\noindent For the NFW mass estimates presented in \S4, we will use
this relation between mass and concentration, thus assuming we can
describe the cluster mass distribution with a single parameter.  We
note that the simulations show considerable scatter in the profiles
from halo to halo. For instance, the values of $c$ for halos of a
given mass have a lognormal dispersion of approximately 0.14 around
the median.

By definition, the virial mass and radius are related through

\begin{equation}
M_{\rm vir}=\frac{4\pi}{3} \Delta_{\rm vir}(z)\rho_{\rm bg}(z)r_{\rm vir}^3,
\end{equation}

\noindent where $\rho_{\rm bg}=3H_0^2\Omega_m(1+z)^3/(8\pi G)$ is the
mean density at the cluster redshift and the virial overdensity
$\Delta_{\rm vir}\approx (18\pi^2+82\xi-39\xi^2)/\Omega(z)$, with
$\xi=\Omega(z)-1$ (Bryan \& Norman 1998). For the $\Lambda$CDM
cosmology considered here, $\Delta_{\rm vir}(0)=337$. 

Cluster masses are also often quoted in terms of $M_\Delta$, which is
the mass contained within the radius $r_{\Delta}$, where the mean mass
density of the halo is equal to $\Delta$ times the critical density at
the redshift of the cluster. For reference with other work, we list
results for a number of commonly used values for $\Delta$. Note
that $M_{200}$ is often referred to as the virial mass, but that our
definition for $M_{vir}$ is different.

The expressions for the tangential shear and surface density for the
NFW profile have been derived by Bartelmann (1996) and Wright \&
Brainerd (2000) and we refer the interested reader to these papers for
the relevant equations.

\subsection{Aperture mass}

Another approach to determine the cluster mass is known as
aperture mass densitometry. It uses the fact that the shear can be
related directly to a density contrast. We use the statistic of Clowe
et al (1998), which is related to the $\zeta$-statistic of Fahlman et
al. (1994). 

In terms of the dimensionless surface density $\zeta_c(r_1)$ is equal
to the mean surface density interior to $r_1$ relative to the mean
surface density in an annulus from $r_2$ to $r_{\rm max}$

\begin{equation}
\zeta_c(r_1)=\bar\kappa(r'<r_1)-\bar\kappa(r_2<r'<r_{\rm max}).
\end{equation}

\noindent It is related to the (observed) shear through

\begin{equation}
\zeta_c(r_1)=2\int_{r_1}^{r_2}d\ln r\langle\gamma_t\rangle+
\frac{2r_{\rm max}^2}{r_{\rm max}^2-r_2^2} \int_{r_2}^{r_{\rm max}}
d\ln r \langle\gamma_t\rangle.
\end{equation}

Equation~8 shows that we can determine the average surface density
within a given aperture up to a constant. If we ignore the surface
density in the annulus, $\zeta_c$ provides a lower limit to the mass.
This demonstrates the usefulness of wide field imaging data: at large
radii, the mean surface density in the annulus should be small and its
contribution relative to the mean surface density within $r_1$
decreases. 

We use the surface density from the NFW model to convert the observed
signal into an estimate of the shear. In this case the mass estimate
is based on the signal measured at large radii, and the correction is
small.

For the mass estimates presented in \S4, we adopt $r_2=600''$ and
$r_{\rm max}=1000''$. We estimate the mean surface density in the
annulus based on the best fit NFW model. Thanks to our ability to
measure the lensing signal out to large radii, the (model dependent)
correction to the mass is only $\sim 10\%$.

\subsection{Complications in weak lensing mass estimates}

Hoekstra et al. (2000) lists various effects which complicate a simple
interpretation of the lensing signal. Some of these have typically
been included in published weak lensing studies, but others are
often ignored. We therefore briefly review them here.

First of all, the observed lensing signal is invariant under the
transformation $\kappa'=(1-\lambda)\kappa+\lambda$, which is usually
referred to as the mass sheet degeneracy (Gorenstein, Shapiro \& Falco
1988): we can determine the surface density up to a constant. This can
also be seen from Eqn.~8, where one needs to estimate the mean surface
density in an annulus to derive the enclosed mass. At very large
radii, however, the cluster surface density should be small, and the
mass sheet degeneracy is less relevant. As mentioned above, we
estimate the mass in the annulus from a mass model, and find that the
correction to the enclosed mass is typically less than $10\%$. Hence,
the uncertainty in the mass due to the mass sheet degeneracy is a few
percent at most for the wide field imaging data used here.

The images of the distant galaxies are not only distorted, but also
magnified. The flux is increased by a factor
$\mu=((1-\kappa)^2-\gamma^2)^{-1}$. This changes the source redshift
distribution as intrinsically fainter galaxies are included as
sources. This effect is small, even in the inner regions of the
cluster where the surface density is large. It is therefore safe to
ignore this effect in our analysis.

A complication which is relevant for our measurements, is the fact
that we do not measure the shear $\gamma$ from the data, but the
distortion (or reduced shear) $g=\gamma/(1-\kappa)$. In the weak
lensing limit $(\kappa\ll 1)$, the distortion is equal to the shear,
but even at the large radii probed here, the amplitude of this effect
is a few percent. Hence, it is important to account for the
$(1-\kappa)$ factor in our cluster mass estimates.

\begin{figure}
\begin{center}
\leavevmode
\hbox{%
\epsfxsize=8.5cm
\epsffile{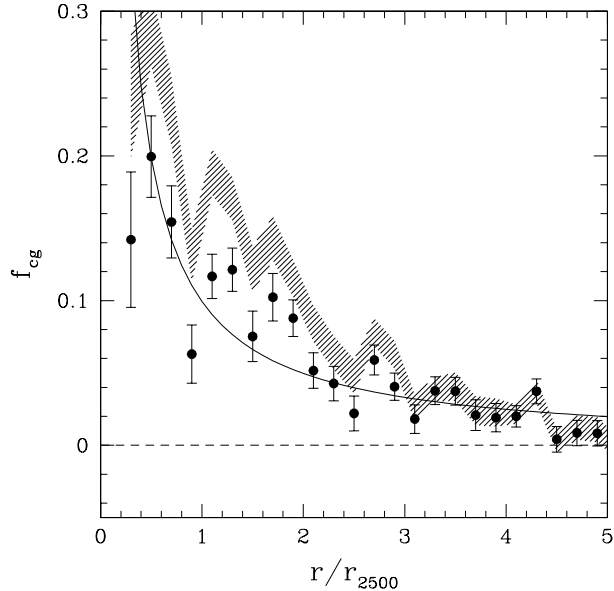}}
\caption{Fraction of cluster galaxies $f_{cg}$ as a function of radius
from the cluster center in units of $r_{2500}$. The points indicate
the contamination after removal of galaxies with colors along the
cluster red-sequence. The shaded are corresponds to the level of
contamination if no color selection is applied. The level of
contamination is reduced by only $\sim 30$\% by applying a color cut. 
However, it is clear that some contamination remains at small
radii. The solid line is the best fit $1/r$ model.}
\label{contam}
\end{center}
\end{figure}

\subsection{Contamination by cluster galaxies}

Weak lensing observations typically lack redshift information for the
galaxies used in this analysis. As a result we cannot distinguish
between unlensed cluster galaxies and lensed sources. The
contamination by cluster galaxies will lower the lensing signal (where
we assume their orientations are random) and therefore needs to be
accounted for, i.e., we need to increase the observed shear by a
factor $1+f_{cg}(r)$, where $f_{cg}$ is the fraction of cluster
galaxies at that radius. Note that foreground field galaxies also
dillute the lensing signal, but this is naturally accounted for in the
calculation of the critical surface density (see Eqn.~2 and the
discussion in \S3.8)

We used our data to examine the excess counts of galaxies as a
function of distance from the cluster. We combined the measurements of
the clusters in our sample. The amount of contamination at a given
angular or physical scale depends on the mass (or richness) of the
cluster. Although the range in mass is relatively small, we found that
the profiles match best if we use the radius in units of $r_{2500}$
(or equivalently the virial radius). Note, that this does not require
a scaling of the counts themselves.

We also examined whether the contamination depends on cluster
redshift.  One might expect this, because the red-sequence occupies
different regions in the color-magnitude diagram as a function of
redshift.  However, we did not find a significant trend with redshift
and we therefore assume the correction is redshift independent.

The results are presented in Figure~\ref{contam}. The shaded area
indicates the fraction of cluster galaxies as a function of radius if
no color selection is used for the sources. At $r_{2500}$ (which
roughly corresponds to $\sim 2$ arcminutes for the clusters considered
here) we find that about 14\% of the galaxies are cluster
members. Thanks to our color information, we can define a cluster
red-sequence and reject galaxies that lie on this sequence. To do so
we identify the red ridge of the cluster color-magnitude
relation and select all galaxies that are up to $\sim 0.3$ magnitudes
bluer. This removes the bright elliptical galaxies quite effectively,
but at fainter magnitudes, the red-sequence is not well defined and
many cluster members are actually blue. Consequently the color
selection reduces the contamination by a modest $\sim 30$\% to 10\% at
$r_{2500}$. This remaining contamination is typically ignored in the
literature.

Comparison to the overdensity of red galaxies shows that the observed
contamination traces the distribution of red galaxies well. At large
radii the measurements are noisy and suffer from the fact that it is
difficult to determine the background level. We used the counts in an
annulus from 10 to 15 arcminutes (more than 5 times $r_{2500}$ for
most clusters) from the cluster centre. This may not be sufficiently
far out, and as a result we may underestimate the contamination at
large radii. It is clear, however, that the contamination is small at
large radii.

To estimate the level of contamination as a function of radius we
assume that $f_{cg}\propto r^{-1}$ (as the data do not allow a good
estimate of the slope). The best fit result is indicated by the solid
line in Figure~\ref{contam}. We use this model to correct the
tangential shear measurements for contamination by cluster members.
The corrections are small, and we determine $r_{2500}$ from the
uncorrected shear profile, and use this to correct the profile and
redetermine $r_{2500}$. This iteration scheme converges rapidly.  We
find that our correction for the residual contamination by cluster
members increases the measured Einstein radii by $\sim 7$\% (see
Table~\ref{tabsis}). The aperture masses listed in Table~\ref{tabmass}
are affected less, and the mass within an aperture of radius of $0.5
h^{-1}$Mpc increases by $\sim 4$\%.

\subsection{Source redshift distribution}

As mentioned above, the interpretation of the lensing signal requires
knowledge of the redshifts of the source galaxies. Based on our data
alone, we do not know the redshifts of the individual sources. The
observed lensing signal, however, is an ensemble average of many
different galaxies, each with their own redshift. As a result, it is
sufficient to know the source redshift distribution to compute
$\langle\beta\rangle$.

The source galaxies are typically too faint to be included in
spectroscopic redshift surveys, although much progress is expected in
the coming years. Instead we use the photometric redshift
distributions determined from the Hubble Deep Fields (Fernandez-Soto
et al. 1999). The clusters in our sample are at relatively low
redshifts, which reduces the uncertainty in the mass measurements
caused by errors in the redshift distribution. The HDF redshift
distribution matches redshift distributions from other (shallower)
photometric redshift surveys, such as COMBO-17 (Wolf et al. 2004) and
the Red-Sequence Cluster Survey (P.~Hsieh private communication).

As discussed in the previous section, we identify the cluster
color-magnitude relation and remove galaxies with similar colors as
the cluster galaxies from the source catalogs. This increases the
lensing signal in the central regions. Consequently the color cut
changes the redshift distribution of the source galaxies, but we find
that the effect is very small. It is negligible for the higher
redshift clusters, but lowers the value for $\langle\beta\rangle$ by
$\sim 2\%$ for clusters at $z\sim 0.2$.

The noise in the shape measurements for the faintest galaxies is
large, which is further amplified by the large correction for
seeing. We therefore limit the range in apparent magnitude for the
sources used in the analysis. The adopted range is listed in
Table~\ref{tabsis}.  Following Hoekstra et al. (2000) we weight the
shapes proportional to the inverse square of the measurement error and
adjust the redshift distribution accordingly. All the considerations
listed above lead to an effective value for $\langle\beta\rangle$,
which is listed in Table~\ref{tabsis}. We estimate a statistical error
in $\langle\beta\rangle\sim 0.01$. This value is similar for all the
clusters in our sample (note, however, that the relative error is
larger for the higher redshift clusters).

When computing the ensemble averaged distortion, one uses an average
value of $\beta$ for the sources. In doing so, one effectively assumes
that the redshift distribution can be approximated by a sheet of
galaxies at a redshift corresponding to the mean value of $\beta$.
As shown by Hoekstra et al. (2000) this results in an overestimate
of the shear by a factor

\begin{equation}
1+\left[{\frac{\langle\beta^2\rangle}{\langle\beta\rangle^2}-1}\right]\kappa.
\end{equation}

For high redshift clusters this effect can be of comparable size as
the correction for the fact that we measure the reduced shear. It is,
however, typically ignored in the literature. We computed the values
for $\langle\beta^2\rangle$ using the HDF photometric redshift
distributions and the results are listed in Table~\ref{tabsis}. We
include this correction for the mass estimates prestented in \S4.

\subsection{Projection effects}

The fact that lensing is sensitive to all matter along the line of
sight complicates the direct comparison with other mass estimates.
One complication is the three-dimensional structure of clusters (e.g.,
Corless \& King, 2006). Although the measurement of the weak lensing
signal does not require any assumptions about the geometry of the
cluster, one does need to make such assumptions when comparing to other
mass indicators, which depend on the cluster mass distribution in a
different way.  Therefore some of the scatter between the various mass
estimates presented in \S4 will be caused by this effect.

Large scale structure gives rise to two distinct contributions, both
of which have been studied in detail. Structures associated with the
cluster, such as filaments, have been studied by Metzler et al.
(1999; 2001) using numerical simulations. Unfortunately, these studies
focussed on the use of aperture mass measurements at large radii,
which makes it difficult to estimate the effect for our mass model
fits. Nevertheless, it is clear from Metzler et al. (2001), that the
effect is significantly reduced by focussing on the central regions of
the cluster, which are much denser than the cosmic web. Alternatively,
provided photometric redshift of the sources are available, one can
reconstruct the three-dimensional mass distribution (e.g., Taylor et
al. 2004) and correct for additional structures along the line of sight.

The other contribution, arises from distant (uncorrelated) large scale
structure. The observed aperture mass $M_{\rm obs}(\theta)$ is the sum
of the actual mass of the cluster $M_{\rm cl}(\theta)$ and the
contribution from all other structure along the line of sight $M_{\rm
LSS}(\theta)$. The expectation value for the latter contribution
vanishes, although the distribution is slightly skewed. Therefore, on
average, the distant large scale structure does not bias the lensing
mass, but it introduces an additional uncertainty in the mass
measurement of size $\langle M_{\rm LSS}^2\rangle^{1/2}$. This effect
was first studied in detail in Hoekstra (2001) and Hoekstra (2003)
using a semi-analytic approach which has been verified using numerical
simulations by White et al. (2002).

The noise introduced by the large scale structure can be written as an
integral of the projected convergence power spectrum $P_\kappa(l)$
multiplied by a filter function $g(l,\theta)$, which depends on the
adopted statistic (Hoekstra 2001)

\begin{equation}
\langle M_{\rm lss}^2\rangle (\theta)=2\pi \int_0^{\infty}
dl~l P_\kappa(l) g(l,\theta)^2.
\end{equation}

The expressions for $P_\kappa(l)$ and $g(l,\theta)$ can be found in
Hoekstra (2001). For the purpose of this paper it suffices to note
that the amplitude of the contribution of distant large scale
structure depends on the adopted cosmology and source redshift
distribution. For the former we adopt a $\Lambda$CDM cosmology with
$\sigma_8=0.85$, while the source redshift distribution is the same as
the one used in the cluster weak lensing analysis.

The resulting additional uncertainty in the value of the Einstein
radius is listed in column~9 in Table~\ref{tabsis}. This additional
error should be added in quadrature to the statistical error (as the
two sources of uncertainty are uncorrelated), resulting in the
total error listed in column~10. Including the contribution from
large scale structure increases the uncertainty in the determination
of the Einstein radius by $10-15\%$.

As shown in Hoekstra (2001) the effect of distant large scale is more
important for aperture mass measurements, which is not surprising,
given that such a mass determination depends on the lensing signal at
large radii (e.g., see Eqn.~9). We find that for the aperture mass
measurements, the contribution from large scale structure increases
the true uncertainty by $20-30\%$ over the statistical error.
Similarly we follow Hoekstra (2003) to estimate the errors for the NFW
model fits. We do not list the additional uncertainties separately,
but note that the large scale structure contribution is included in
the errors listed in Table~\ref{tabmass}.

\begin{figure}
\begin{center}
\leavevmode
\hbox{%
\epsfxsize=8cm
\epsffile{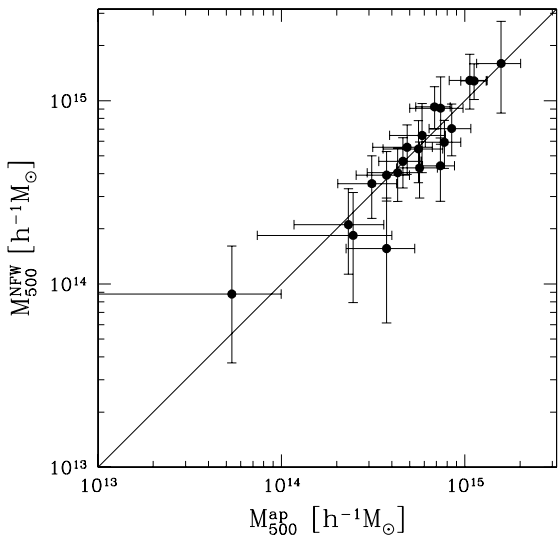}}
\caption{Comparison of $M_{500}$ determined from an NFW fit to the
data at radii $0.25<r<1h^{-1}$ Mpc and the value for $M_{500}$ derived
from the aperture mass method. The latter predominantly uses data at
large radii, whereas the former is based on the lensing signal at
small scales (note that the fitted range is smaller than for the
values listed in Table~\ref{tabmass}. The two measurements are almost
independent as the average $r_{500}$ for the sample studied here is
$870h^{-1}$kpc. The solid line corresponds to the line of equality. }
\label{m500_comp}
\end{center}
\end{figure}

\begin{table*}
\caption{Weak lensing mass estimates \label{tabmass}}
\centering
\begin{tabular}{lcccccccc}
\hline
\hline
(1)        & (2)         &  (3)        & (4) & (5) & (6) & (7) & (8) & (9)\\
name       & $\sigma_{\rm WL}$ & $M(<0.5h^{-1}{\rm Mpc})$  & $M_{2500}$ & $M_{500}$ & 
$M^{\rm NFW}_{2500}$ & $M^{\rm NFW}_{500}$ & $M^{\rm NFW}_{200}$ & $M^{\rm NFW}_{\rm vir}$ \\
           & [km/s]      &     &       &     &     &     &        \\ 
\hline
A2390      & $1117^{+76}_{-82}$   & $5.2\pm 0.6$ & $2.4\pm 0.5$ & $6.8\pm  1.5$ & $2.9\pm 0.6$        & $9.2^{+2.0}_{-1.9}$  & $14.6^{+3.1}_{-2.9}$ & $18.0^{+3.8}_{-3.6}$  \\
MS 0016+16 & $1164^{+151}_{-173}$ & $7.9\pm 1.1$ & $3.2\pm 0.7$ & $15.8\pm 4.3$ & $4.2^{+1.4}_{-1.3}$ & $16.0^{+5.3}_{-4.9}$ & $27.0^{+9.0}_{-8.4}$ & $32.0^{+10.7}_{-9.9}$ \\
MS 0906+11 & $880^{+99}_{-111}$   & $3.7\pm 0.7$ & $1.6\pm 0.4$ & $7.4\pm  1.5$ & $1.5\pm0.4$         & $4.4^{+1.2}_{-0.2}$  & $6.7^{+1.8}_{-1.8}$  & $8.3^{+2.3}_{-2.3}$   \\
MS 1224+20 & $837^{+133}_{-158}$  & $3.0\pm 0.9$ & $1.0\pm 0.4$ & $2.3\pm  1.2$ & $0.8\pm0.4$         & $2.1^{+1.2}_{-1.0}$  & $3.2^{+1.8}_{-1.5}$  & $3.8^{+2.1}_{-1.8}$   \\
MS 1231+15 & $566^{+145}_{-195}$  & $0.8\pm 0.6$ & $0.4\pm 0.2$ & $0.5\pm  0.5$ & $0.4\pm0.2$         & $0.9^{+0.5}_{-0.5}$  & $1.3^{+0.7}_{-0.7}$  & $1.5^{+0.9}_{-0.9}$   \\
MS 1358+62 & $1048^{+102}_{-113}$ & $4.3\pm 0.8$ & $1.8\pm 0.4$ & $5.6\pm  2.0$ & $1.8^{+0.6}_{-0.5}$ & $5.5^{+1.9}_{-1.7}$  & $8.5^{+3.0}_{-2.6}$  & $10.3^{+3.6}_{-3.1}$   \\
MS 1455+22 & $964^{+87}_{-95}$    & $3.3\pm 0.7$ & $1.2\pm 0.3$ & $3.7\pm  1.2$ & $1.4^{+0.4}_{-0.3}$ & $3.9^{+1.0}_{-1.0}$  & $6.0^{+1.6}_{-1.5}$  & $7.3^{+1.9}_{-1.8}$   \\
MS 1512+36 & $722^{+145}_{-181}$  & $2.1\pm 0.8$ & $0.6\pm 0.3$ & $2.5\pm  1.6$ & $0.7^{+0.4}_{-0.3}$ & $1.8^{+1.1}_{-0.9}$  & $2.8^{+1.6}_{-1.4}$  & $3.3^{+1.9}_{-1.6}$   \\
MS 1621+26 & $998^{+128}_{-146}$  & $5.3\pm 1.0$ & $1.7\pm 0.7$ & $5.8\pm  2.0$ & $2.0^{+0.8}_{-0.6}$ & $6.5^{+2.8}_{-1.9}$  & $10.3^{+4.4}_{-3.0}$ & $12.3^{+5.2}_{-3.6}$  \\
\hline
A68        & $1036^{+89}_{-97}$   & $4.4\pm 0.8$ & $1.9\pm 0.4$ & $4.8\pm  1.8$ & $1.8^{+0.5}_{-0.4}$ & $5.6^{+1.6}_{-1.3}$  & $8.6^{+2.5}_{-2.1}$  & $10.5^{+3.1}_{-2.5}$   \\
A209       & $898^{+92}_{-102}$   & $3.9\pm 0.9$ & $1.5\pm 0.6$ & $5.7\pm  1.4$ & $1.5^{+0.5}_{-0.4}$ & $4.3^{+1.4}_{-1.1}$  & $6.6^{+2.1}_{-1.7}$  & $8.0^{+2.5}_{-2.1}$   \\
A267       & $1008^{+90}_{-99}$   & $3.3\pm 0.6$ & $1.5\pm 0.3$ & $4.3\pm  1.4$ & $1.4\pm0.4$         & $4.0^{+1.2}_{-1.2}$  & $6.2^{+1.9}_{-1.8}$  & $7.5^{+2.3}_{-2.2}$   \\
A383       & $701^{+138}_{-171}$  & $2.6\pm 0.7$ & $0.6\pm 0.3$ & $3.7\pm  1.6$ & $0.6^{+0.4}_{-0.3}$ & $1.6^{+0.9}_{-0.8}$  & $2.3^{+1.3}_{-1.2}$  & $2.8^{+1.6}_{-1.5}$   \\
A963       & $844^{+99}_{-112}$   & $2.7\pm 0.6$ & $1.0\pm 0.3$ & $3.1\pm  1.1$ & $1.3\pm 0.4$        & $3.5^{+1.1}_{-1.0}$  & $5.3^{+1.6}_{-1.5}$  & $6.5^{+2.0}_{-1.9}$   \\
A1689      & $1370^{+65}_{-68}$   & $6.7\pm 0.7$ & $3.7\pm 0.5$ & $11.2\pm 1.8$ & $4.0^{+0.8}_{-0.7}$ & $12.8^{+2.7}_{-2.3}$ & $20.4^{+4.2}_{-3.6}$ & $25.5^{+5.3}_{-4.5}$  \\
A1763      & $1060^{+87}_{-95}$   & $4.9\pm 0.7$ & $2.3\pm 0.4$ & $8.5\pm  2.2$ & $2.3\pm 0.6$        & $7.0^{+2.0}_{-1.7}$  & $11.0^{+3.0}_{-2.7}$ & $13.5^{+3.7}_{-3.3}$  \\
A2218      & $1042^{+87}_{-94}$   & $4.5\pm 0.7$ & $2.0\pm 0.5$ & $4.6\pm  1.2$ & $1.6^{+0.5}_{-0.4}$ & $4.7^{+1.5}_{-1.3}$  & $7.1^{+2.3}_{-1.9}$  & $8.8^{+2.8}_{-2.4}$   \\
A2219      & $1074^{+82}_{-89}$   & $5.0\pm 0.7$ & $2.3\pm 0.4$ & $7.7\pm  1.7$ & $2.0^{+0.6}_{-0.5}$ & $5.9^{+1.7}_{-1.4}$  & $9.2^{+2.7}_{-2.2}$  & $11.3^{+3.2}_{-2.7}$  \\
\hline
A370       & $1359^{+90}_{-96}$   & $6.5\pm 0.9$ & $3.1\pm 0.5$ & $10.6\pm 2.5$ & $3.7^{+1.1}_{-0.9}$ & $12.9^{+3.8}_{-3.2}$ & $21.1^{+6.2}_{-5.3}$ & $25.5^{+7.5}_{-6.4}$  \\
CL0024+16  & $1140^{+111}_{-123}$ & $5.5\pm 0.9$ & $2.3\pm 0.5$ & $7.4\pm  2.4$ & $2.7^{+0.9}_{-0.8}$ & $9.1^{+3.2}_{-2.7}$  & $14.7^{+5.1}_{-4.4}$ & $17.6^{+6.2}_{-5.3}$  \\
\hline
\hline
\end{tabular}
\bigskip

\begin{minipage}{0.9\linewidth}

{\footnotesize Column 1: cluster name; Column 2: velocity dispersion
derived from the SIS model; Column 3: projected mass enclosed in an
aperture of radius $0.5 h^{-1}$ Mpc in units of
$10^{14}h^{-1}M_\odot$; Column 4,5: the mass from the aperture mass
measurement. The values correspond to the mass in a sphere of radius
$r_{2500}$ and $r_{500}$, where the mean density is 2500 and 500 times
the critical density at the redshift of the cluster, respectively;
Columns 6-9: $M_{2500}$, $M_{500}$, $M_{200}$ and the virial mass
$M_{\rm vir}$ as derived from an NFW fit to the tangential distortion
at radii $0.25-1.5h^{-1}$Mpc; All masses are in units of
$10^{14}h^{-1}M_\odot$ and the error bars correspond to the 68\%
confidence interval. We refer the reader to \S3.4 for the precise
definitions of the virial mass and $M_\Delta$ employed in this
paper.}
\end{minipage}
\end{table*}

\section{Mass estimates}

Table~\ref{tabmass} lists the various weak lensing mass estimates.
Column 2 shows the inferred line-of-sight velocity dispersions
from the SIS model fit discussed in \S3.3. These results can
be compared directly to velocity dispersions of galaxies from
spectroscopic observations, which is done in \S4.2. The error bars
include both the statistical and distant large scale structure
errors. 

Column~3 shows the projected mass within an aperture of $0.5h^{-1}$
Mpc using the aperture mass technique. This measurement requires no
assumptions about the geometry or mass profile (apart from the small
correction for the mean surface density in the annulus). However, it
is difficult to compare this result to other mass indicators such as
X-ray properties because the latter typically assume spherical
symmetry.

The virial mass as defined by Eqn.~7 is of interest, because it has
some physical meaning. Given the low density in the $\Lambda$CDM
model, the virial radius is rather large for the clusters considered
here, and cannot be computed from the aperture mass method for two
reasons. The first is that the area covered by the observations is not
large enough. The second reason is more fundamental: as described in
the previous section, projection effects limits the accuracy of the
weak lensing mass determination.  At the virial radius these errors
are too large for a useful measurement.

Instead, we list the values for $M_{2500}$ and $M_{500}$ in
Table~\ref{tabmass}, where the overdensities are measured with respect
to the critical density at the redshift of the cluster. For these
values of $\Delta$ we can derive reasonably accurate masses in a model
independent manner. For the comparison with the ASCA data in \S4.3,
the $M_{2500}$ is most relevant as $r_{2500}$ is close to the radius
out to which X-ray temperatures are measured.

We also fitted the NFW profile to the lensing signal at radii
$0.25<r<1.5 h^{-1}$ Mpc, which is the same range as was used for the
SIS model fit. The resulting virial masses for this model are
presented in Column~9 of Table~\ref{tabmass}. For completeness, we
also list the corresponding masses for a range of other values of
$\Delta$ which are commonly found in the literature (see \S3.4 for
details).

We can measure $M_{500}$ in two different ways, which are almost
independent. The aperture mass is predominantly based on data at large
radii, whereas the NFW model fit uses measurements at small
scales. The comparison of these mass estimates provides an additional
check of the weak lensing analysis.  We compare these two mass
estimates in Figure~\ref{m500_comp}, where we limited the NFW fit to
radii $0.25<r<1 h^{-1}$ Mpc, to minimize the overlap with the scales
used for the aperture mass measurement (the mean value for $r_{500}$
for the clusters in our sample is $870h^{-1}$kpc). We find that the
mass estimates agree quite well.

\subsection{Comparison with published weak lensing results}

Weak lensing masses have been published for a fair number of the
clusters in our sample, based on observations using a wide range of
telescopes and cameras. In this section we compare our results to the
literature values. Before proceeding, we want to stress that in many
cases the comparison is crude at best, because of differences in the
assumed source redshift distribution, removal of cluster members,
assumed mass model, etc.

The first detection of a weak lensing signal was presented in Tyson et
al. (1991) based on the analysis of A1689. Since then, several mass
estimates for this cluster have been published, and we focus on two
recent ones. Based on wide field imaging data from the ESO 2.2m
telescope, Clowe \& Schneider (2001) estimate a velocity dispersion of
$\sigma=1162\pm40$ km/s (using a mean source redshift of $\sim 0.5$),
which implies a mass which is about 30\% lower than our value. The
analysis presented in Clowe \& Schneider (2001) lacked color
information, and hence a possible explanation for the lower signal
could be contamination by cluster members. The comparison with the
measurement of Bardeau et al. (2005) is of particular interest because
it is based on the same data used here, although the data processing
and weak lensing analysis are completely independent. Bardeau et
al. (2005) list an Einstein radius of $22\pm3$ arcseconds, using a fit
to the lensing signal from 70'' to 1100''. Matching our measurement to
their range, and ignoring contamination by cluster members, lowers our
estimate for the Einstein radius from $32.\pm2.6$ to $25.9\pm1.7$
arcseconds, which is still 18\% higher than Bardeau et
al. (2005). More recently, Limousin et al.  (2006) used a combined
strong and weak lensing analysis. They list a value of
$M_{200}=(19\pm3)\times 10^{14}h^{-1}\msun$, which is in excellent
agreement with our result. On small scales, however, the weak lensing
analysis appears to underestimate the mass.  Limousin et al. (2006)
find a projected mass of $(2.7\pm0.4)\times 10^{14}h^{-1}\msun$ within
a radius of 45\farcs. Extrapolating the weak lensing aperture mass
measurements inwards to this small scale, we find a mass of $(2\pm
0.3)\times 10^{14}h^{-1}\msun$.

Another well studied cluster is MS1224+20, because of its apparent
high mass-to-light ratio. Fahlman et al. (1994) list a projected mass
within a $2\farcm76$ radius of $3.5\times 10^{14}h^{-1}\msun$, which
is only slightly larger than our result of $(3.0\pm0.6)\times
10^{14}h^{-1}\msun$. The cluster was also observed by Fischer (1999)
who inferred a velocity dispersion of $\sim 1300$ km/s, considerably
higher than our estimate. The lensing signal presented in Fischer
(1999) barely decreases with distance, which is not observed in our
data, and we cannot find an obvious explanation for this difference.

Squires et al. (1996a) studied A2390 and list a projected mass
enclosed within a 100'' aperture of $1.8\times 10^{14}h^{-1}\msun$, in
excellent agreement with our estimate of $2.1\times 10^{14}h^{-1}\msun$.
For A2218, Squires et al. (1996b) list a mass $M(<3\farcm
5)=(4.6\pm1.4) \times 10^{14} h^{-1}\msun$, which is higher than our
value of $3.4\times 10^{14}h^{-1}\msun$. 

Dahle et al. (2002) studied a sample of 38 X-ray luminous clusters, 6
of which overlap with our sample. We compared the estimates for the
velocity dispersion listed in Table~2 of Dahle et al. (2002) to our
results and found fair agreement for A209, A267, A963, A1763, and
A2219. However, Dahle et al. (2002) list a velocity dispersion of
$1650\pm220$ for A68, which is significantly higher than our
measurement. Unfortunately we were unable to identify a reason
for this difference.

The mass distribution of MS1358+62 was studied by Hoekstra et
al. (1998) using a mosaic of WFPC2 pointings. Hoekstra et al. (2002)
provide an updated value for the velocity dispersion of
$\sigma=835^{+52}_{-56}$. This number is lower than the value of
$1048^{+102}_{-113}$ km/s found in the CFH12k analysis, but could be
due to differences in the range of scales that were fitted.

CL0024+16 was also studied using HST observations. Kneib et
al. (2003) were able to probe the lensing signal out to $\sim
10'$ by sparsely covering the area with WFPC2 pointings. A direct
comparison is complicated by the fact that Kneib et al. (2003) do not
list the total mass of the cluster, but give the masses of two mass
concentrations instead. The masses are obtained from NFW fits to the
data. Adding the masses yields a value of
$M_{200}=5.7\pm1.3\times10^{14}h^{-1}\msun$. Our data imply a significantly
higher mass of $M_{200}=15^{+5}_{-4}\times10^{14}h^{-1}\msun$.
By considering only two clumps, Kneib et al. (2003) might have
underestimated the mass, as a more diffuse mass distribution would not
have been accounted for. To avoid this problem we also fitted a SIS
model to the tangential distortion presented in Figure~7 of Kneib et
al. (2003).  This yields an Einstein radius of $r_E\sim 18''$.  Kneib
et al. (2003) use sources with $23<I<26$, which yields a value of
$\beta=0.54$.  Hence their signal corresponds to a velocity dispersion
of 1075 km/s, which is in excellent agreement with our findings.

In summary, the agreement between our measurements and the literature
values is fair, but there are a number of cases where the results are
discrepant without a clear cause. However, based on the excellent
quality of the CFH12k data used here, the fact that our weak lensing
technique is well tested, and the comparison with other mass
estimators (see following sections), we are confident that our results
are robust and accurate.

\begin{table}
\centering
\caption{X-ray and dynamical properties\label{tabother}}
\begin{tabular}{lcccc}
\hline
\hline
(1)        & (2)                    & (3)      & (4)  & (5)\\
name       & $L_x$                  & $kT_x$   & $\sigma_{\rm dyn}$   & ref. \\       
          & [$10^{44}h^{-2}$erg/s] & [keV]    & [km/s]               &      \\
\hline
A2390      & 30.5 & $9.2\pm0.6$            &  $1262^{+89}_{-68}$   & 1 \\  
MS 0016+16 & 39.3 & $8.7^{+0.8}_{-0.7}$    &  $1127^{+166}_{-112}$ & 1 \\  
MS 0906+11 & 6.65 & $6.1\pm0.4$            &  $886^{+78}_{-68} $   & 1 \\  
MS 1224+20 & 3.3  & $4.8^{+1.2}_{-1.0}$    &  $831^{+129}_{-57}$   & 1 \\  
MS 1231+15 & $-$  & $-$                    &  $686^{+65}_{-50}$    & 1 \\  
MS 1358+62 & 9.3  & $6.7^{+0.6}_{-0.5}$    &  $1003^{+61}_{-52}$   & 1 \\  
MS 1455+22 & 12.7 & $4.5\pm0.2$            &  $1032^{+130}_{-95}$  & 1 \\  
MS 1512+36 & 4.1  & $3.6^{+0.9}_{-0.7}$    &  $575^{+138}_{-90}$   & 1 \\  
MS 1621+26 & 8.1  & $6.5^{1.3}_{-1.0}$     &  $839^{+67}_{-53}$    & 1 \\  
\hline
A68        & 11.3 & $8.0^{+0.8}_{-0.6}$    & $-$                   &   \\
A209       & $-$  & $-$                    & $-$                   &   \\
A267       & 8.6  & $5.9^{+0.5}_{-0.4}$    & $-$                   &   \\
A383       & $-$  & $-$                    & $-$                   &   \\
A963       & 9.4  & $6.6\pm0.4$            & $-$                   &   \\
A1689      & 25.1 & $9.2^{+0.4}_{-0.3}$    & $1172^{+123}_{-99}$   & 2 \\  
A1763      & 15.6 & $7.7^{+0.5}_{-0.4}$    & $-$                   &   \\  
A2218      & 8.3  & $7.0^{+0.4}_{-0.3}$    & $1222^{+147}_{-109}$  & 2 \\  
A2219      & 32.8 & $9.8^{+0.7}_{-0.6}$    & $-$                   &   \\
\hline
A370       & 13.4 & $7.2\pm0.8$            & $859^{+118}_{-112}$   & 2 \\  
CL0024+16  & 3.5  & $5.2^{+2.0}_{-1.3}$    & $911^{+81}_{-107}$    & 2 \\
\hline
\hline
\end{tabular}

\bigskip

\begin{minipage}{\linewidth}

{\footnotesize Column 1: cluster name; Column 2: Bolometric X-ray
luminosity from Horner (2001).  The values listed here are corrected
for galactic absorption, and transformed to the cosmology adopted in
this paper. Column 3: X-ray temperature from Horner (2001) based on
ASCA observations.  The errors indicate the 90\% confidence
intervals. Column 4: velocity dispersion of cluster galaxies; Column
5: references for velocity dispersions. (1) Borgani et al. (1999); 
(2) Girardi \& Mezzetti (2001)}

\end{minipage}

\end{table}

\begin{figure}
\begin{center}
\leavevmode
\hbox{%
\epsfxsize=8cm
\epsffile{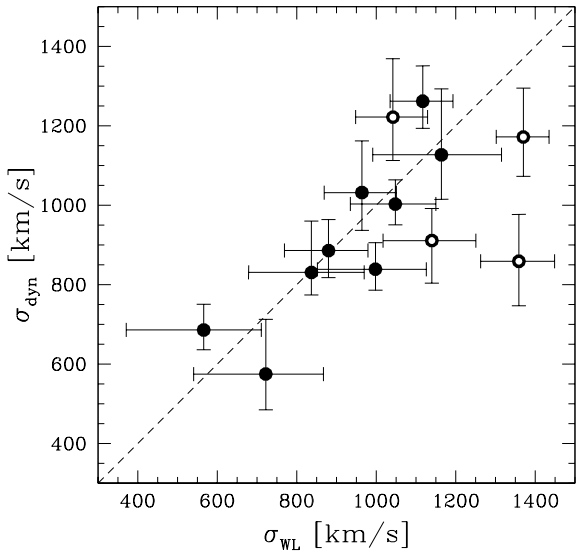}}
\caption{Plot of the velocity dispersion derived from the SIS fit to
the observed tangential distortion versus the velocity dispersion of
cluster galaxies from spectroscopic observations. The dashed line
indicates the line of equality between the two estimates of the
velocity dispersion. The large solid points are the clusters in the
CNOC1 sample, and the open circles correspond to the results presented
in Girardi \& Mezzetti (2001).}
\label{dynmass}
\end{center}
\end{figure}

\subsection{Comparison with dynamical mass estimates}

The measurement of the line-of-sight velocity dispersion clearly
benefits from observing a large number of galaxies.  Such observations
decrease the statistical error, but also allow for a better rejection
of interlopers and for the identification of substructures. These
studies are time consuming, and as a result only a relatively small
number of clusters have been studied in sufficient detail. 

The clusters studied in this paper are well studied clusters and
therefore have relatively good spectroscopic coverage. Nine of the
clusters are part of the CNOC1 redshift survey sample (e.g., Carlberg
et al., 1996; Yee et al., 1996).  Borgani et al. (1999) (re-)derived
velocity dispersions for the clusters in this sample using an
interloper rejection scheme which is more sophisticated than the one
used by Carlberg et al. (1996), which also enabled them to measure
separate velocity dispersions for the two clusters in the field of
MS0906. The velocity dispersions derived by Borgani et al.  (1999) are
listed in Column~4 of Table~\ref{tabother}. The dynamical data for the
remaining clusters in our sample are also listed in
Table~\ref{tabother}.

Figure~\ref{dynmass} shows the velocity dispersion of the cluster
galaxies versus the weak lensing estimate using the SIS fit to the
data. The solid points correspond to the clusters in the CNOC1
sample. The agreement between the weak lensing and dynamical
measurements is excellent for these data. The open points correspond
to measurements from Girardi \& Mezzetti (2001), which also agree with
the lensing estimates, albeit with a larger scatter. Compared to the
CNOC1 sample, these four clusters show evidence of more complicated
dynamics.

\begin{figure}
\begin{center}
\leavevmode
\hbox{%
\epsfxsize=8.5cm
\epsffile{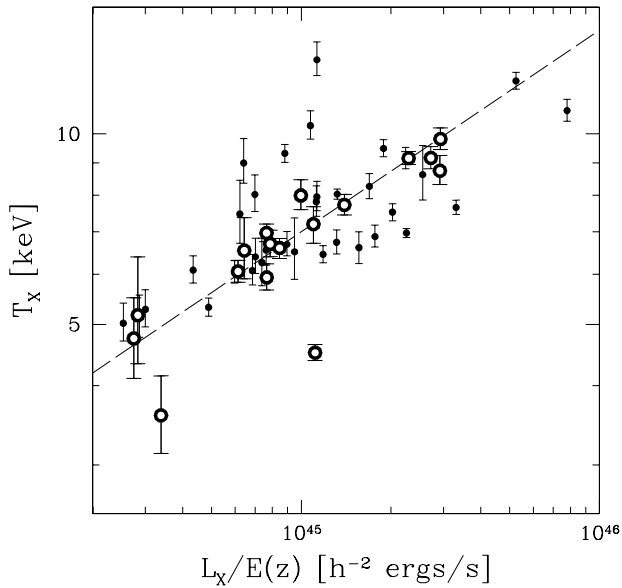}}
\caption{Plot of the ASCA temperature from Horner (2001) as a function
of bolometric X-ray luminosity. The large open points correspond
to the clusters studied in this paper. The small points indicate the
clusters proposed as part of the Canadian Cluster Comparison Project.
Thirty of these have been observed using Megacam on CFHT.
Note the tight $L-T$ relation for the clusters studied in this paper,
which does not appear to be representative of the full population
of galaxy clusters studied as part of CCCP.}
\label{lt}
\end{center}
\end{figure}

\subsection{Comparison with X-ray properties}

Much work has been devoted to the X-ray properties of galaxy
clusters. In particular scaling relations between the X-ray
luminosity, temperature and mass are of great interest as they provide
clues about cluster formation.  For instance, simple self-similar
models (e.g., Kaiser 1986; Bryan \& Norman 1998) predict power law
relations between the mass, temperature and luminosity.  More detailed
numerical simulations (e.g., Evrard, Metzler \& Navarro, 1996; Bryan
\& Norman, 1998) also provide evidence for simple scaling relations.
If the gas is virialised, the mass $M_\Delta$  is given by

\begin{equation}
E(z)M_\Delta \propto T^{3/2}
\end{equation}

\noindent where

\begin{equation}
E(z)= \frac{H(z)}{H_0}=\sqrt{\Omega_m (1+z)^3+\Omega_\Lambda}
\end{equation}

\noindent for flat cosmologies. The relevant temperature is the mean
mass-weighted temperature within the radius $r_\Delta$. As mentioned
earlier, X-ray observations typically measure temperatures out to
$r_{2500}$, which is the radius we will focus on.

The source of the X-ray emission is bremmsstrahlung and therefore one
expects that $L_X/E(z)\propto T^2$. The observed slope is found to be
steeper (e.g., Edge \& Stewart 1991; Markevitch, 1998; Arnaud \&
Evrard, 1999). For clusters with temperatures $\ge 2$keV the slope is
$\sim 3$, with even steeper slopes observed for lower temperature
galaxy groups (e.g., Helsdon \& Ponman 2000). Allen \& Fabian (1998)
have argued that the steeper slope, compared to the self-similar case,
of the hotter clusters is caused by the effects of cool central
components. But studies that attempt to account for this, still find
steeper slopes, with values $\sim 2.7$ (e.g., Markevitch, 1998; Lumb
et al. 2004).

When deriving X-ray temperatures, various groups employ different
approaches to deal with temperature gradients and cool cores. To avoid
introducing scatter caused by variations in the analysis method, we
use the results from Horner (2001). The temperatures are based on ASCA
observations. Although these measurements are not necessarily the most
recent and accurate, they do provide a large sample, analysed
homogeneously. As a caveat, we note that Horner (2001) did not attempt
to correct for the cool centres of clusters. As part of the Canadian
Cluster Comparison
Project\footnote{http://www.astro.uvic.ca/$\sim$hoekstra/CCCP.html}
(CCCP) we will derive X-ray properties in a consistent manner from
modern Chandra and XMM observations, accounting for temperature
variations.

\begin{figure*}
\begin{center}
\leavevmode
\hbox{%
\epsfxsize=8cm
\epsffile{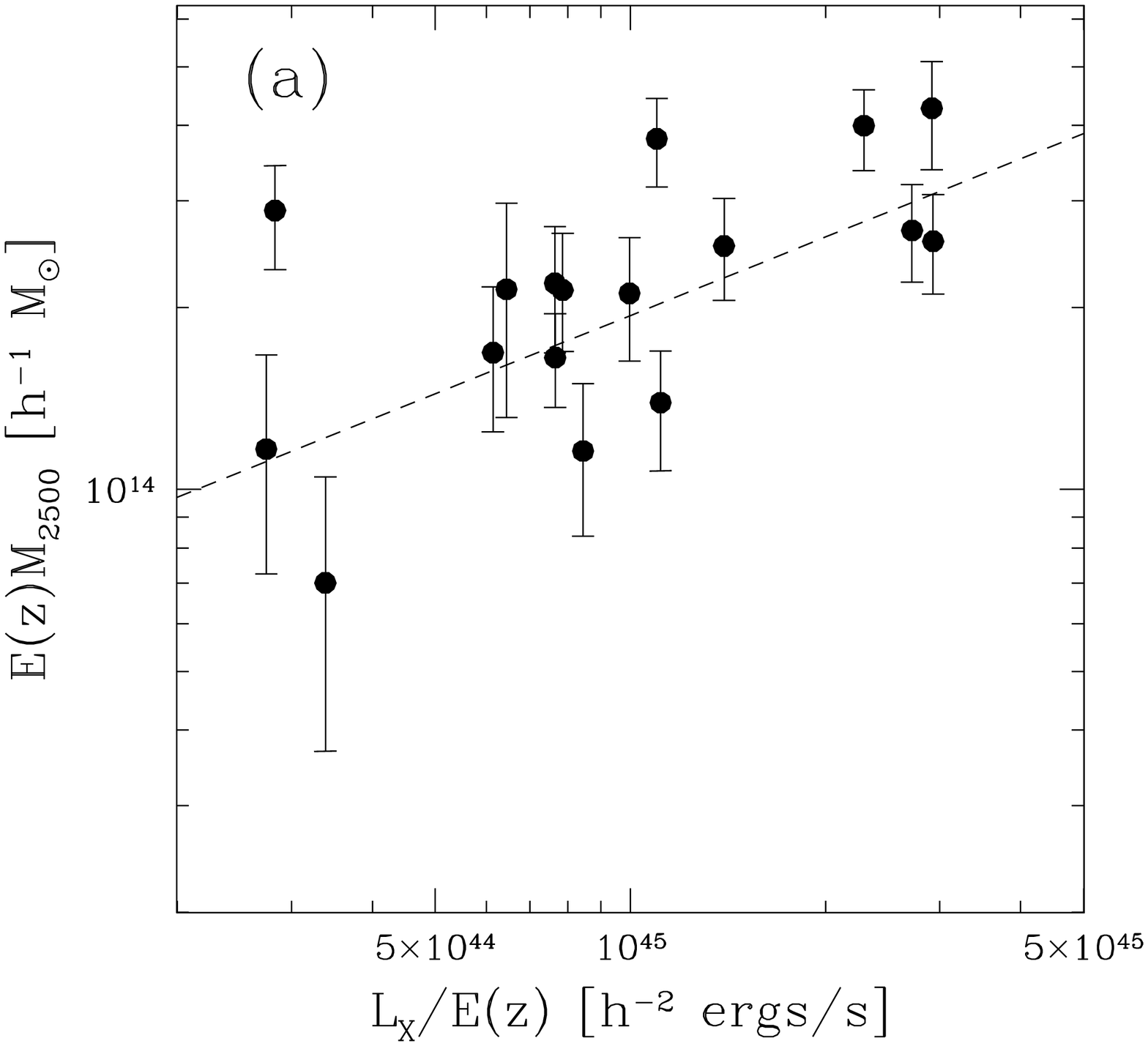}
\epsfxsize=8cm
\epsffile{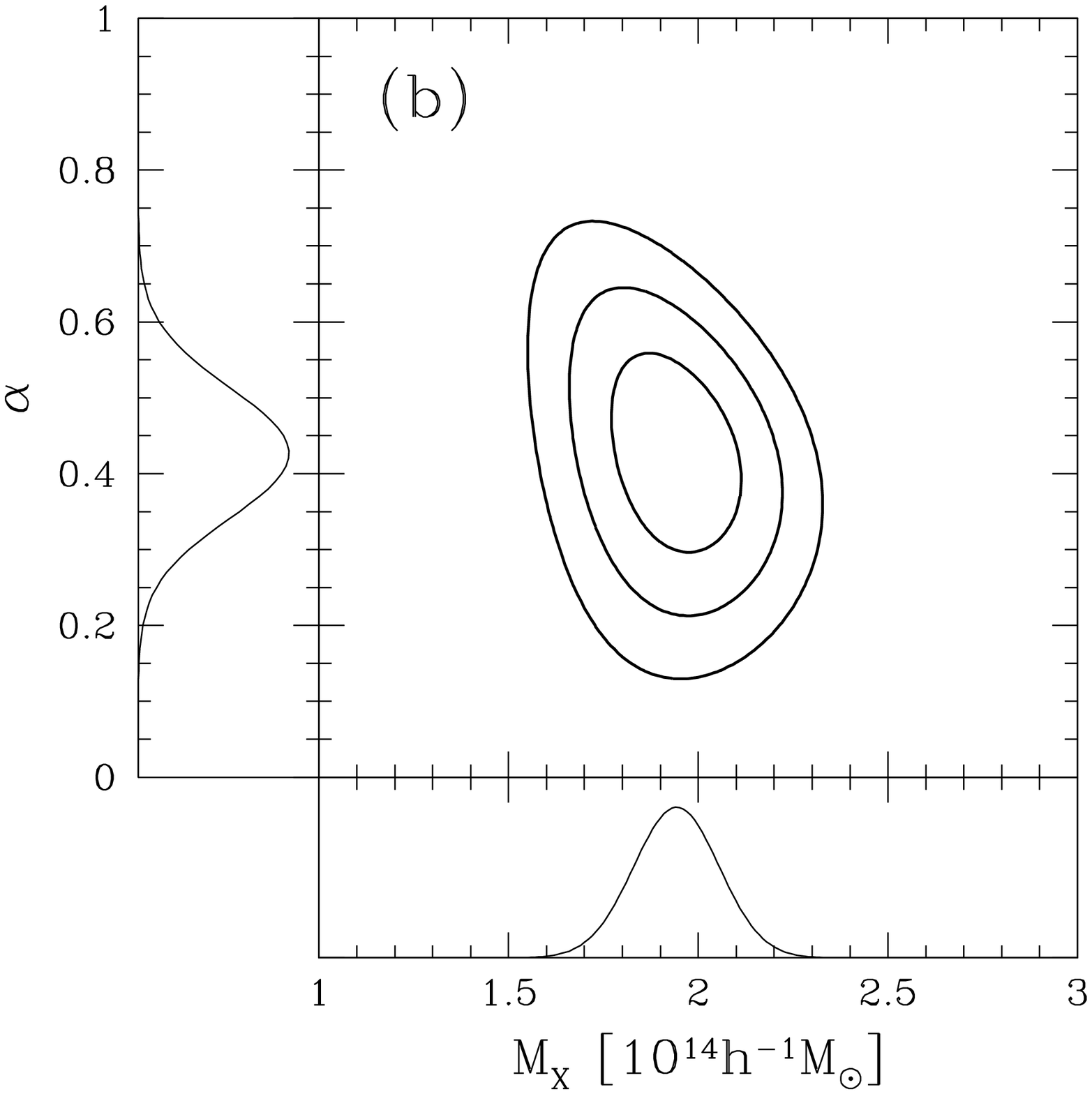}}
\hbox{%
\epsfxsize=8cm
\epsffile{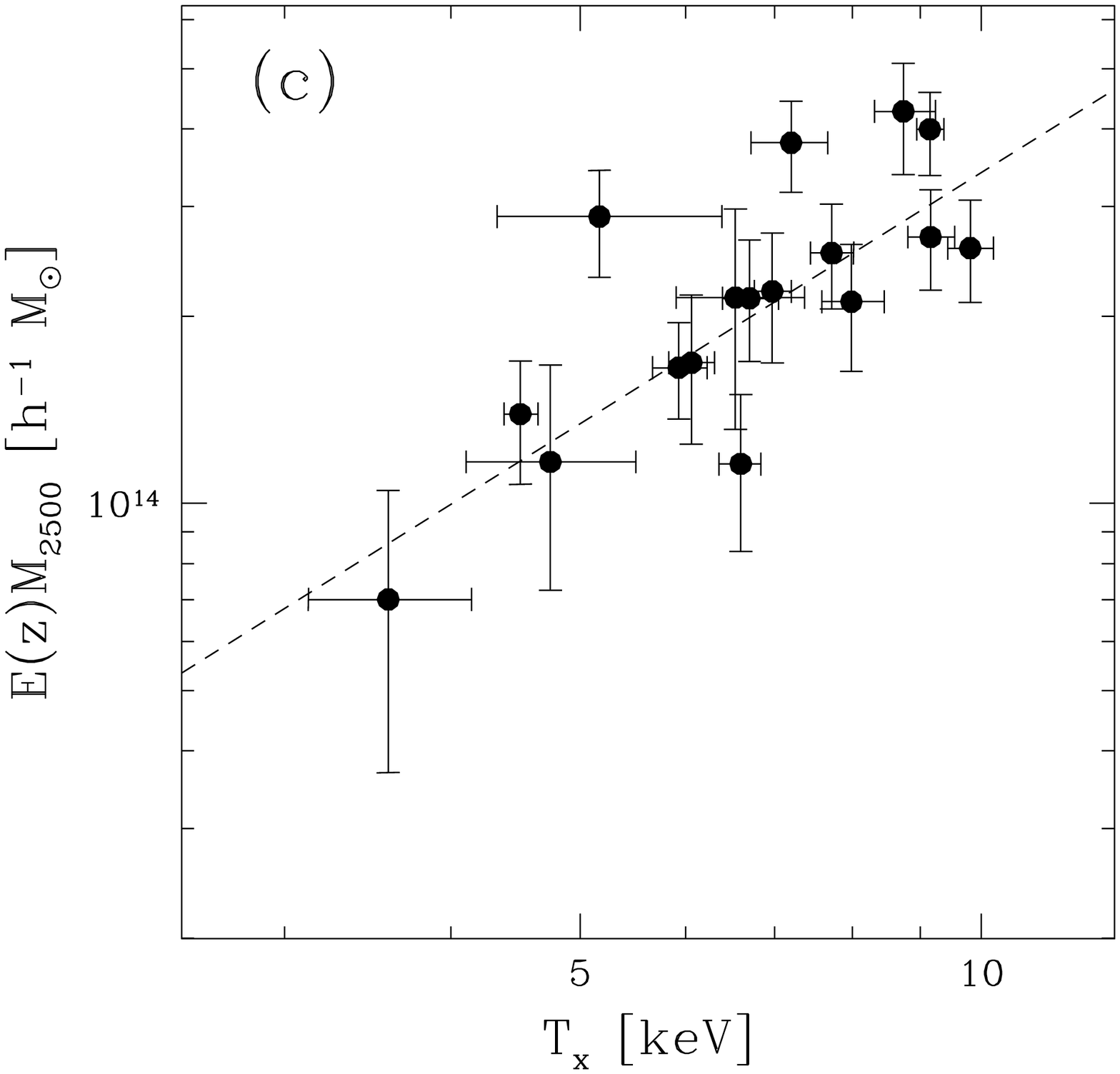}
\epsfxsize=8cm
\epsffile{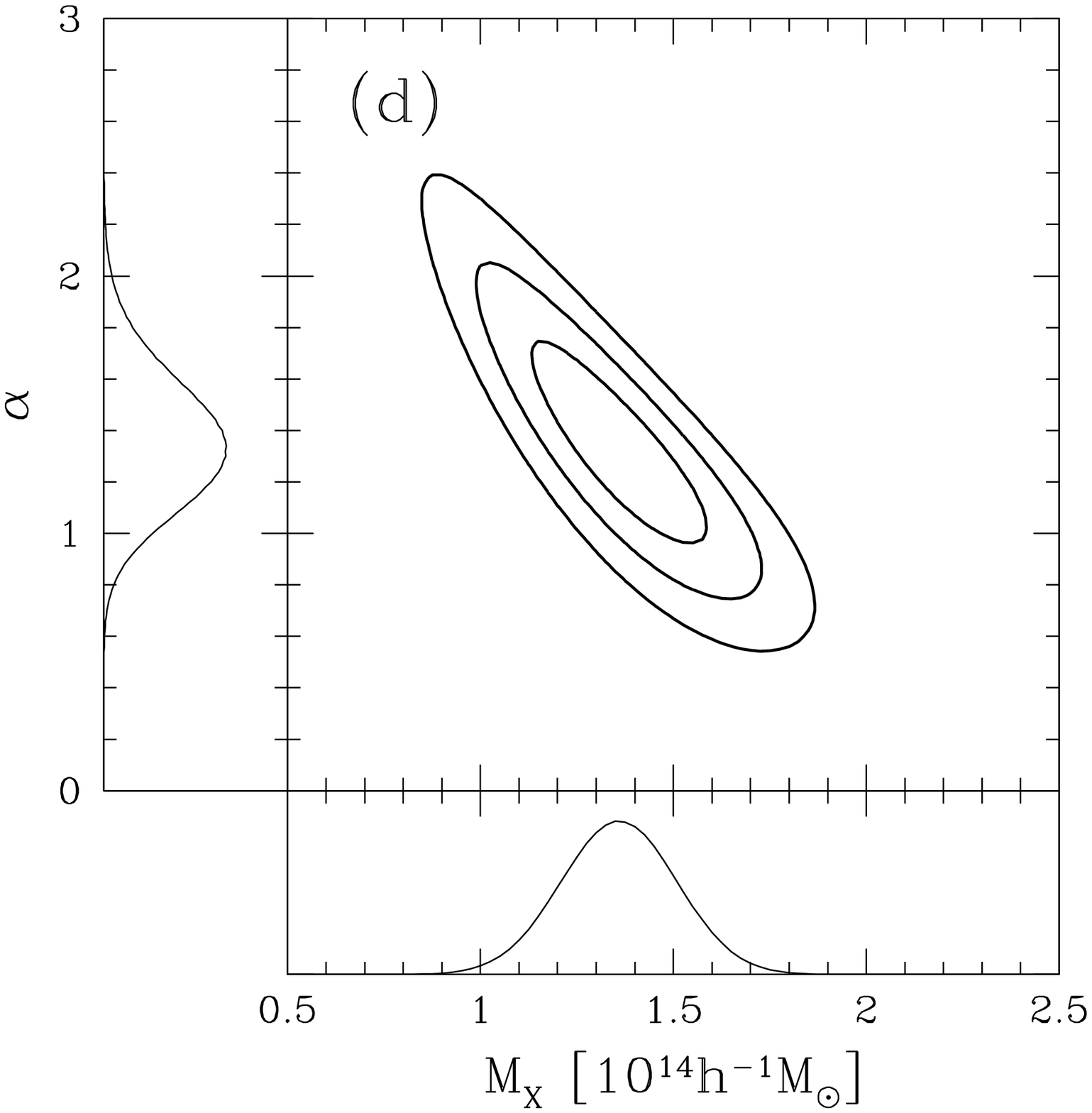}}
\caption{{\it Panel (a):} $M_{2500}$ as a function of the bolometric
X-ray luminosity. To account for the range in redshift of the
clusters, the mass and luminosity have been rescaled using the
corresponding value for $E(z)$. {\it Panel (b):} $M_{2500}$ as a
function of the X-ray temperature. The dashed lines indicate the best
fit power law relations. {\it Panel (a):} Likelihood contours for the slope of the
$M_{2500}-L_X$ relation and the mass of a cluster with a luminosity of
$L_X=10^{45}$ergs/s at $z=0$. {\it Panel (b):} Likelihood contours for
the slope of the $M_{2500}-T_X$ relation and the mass of a cluster
with a temperature of $T_X=5$keV at $z=0$. The contours indicate the
68.3\%, 95.4\% and 99.7\% confidence limits on two parameters jointly.
The side panels show the probability density distribution for each
parameter (while marginalising over the other).}
\label{mass_x}
\end{center}
\end{figure*}

The large open points in Figure~\ref{lt} indicate the temperature as a
function of bolometric luminosity for the clusters studied in this
paper. With the exceptions of MS1455+22 an MS1512+36 (both strong
`cooling flow' clusters), the clusters follow a very tight
relation. If we exclude these outliers, we find a slope of
$3.57\pm0.23$ for our sample.

The aim of the CCCP is not only to derive X-ray temperatures in a
consistent manner, but also involves the systematic study of the
scatter in the relation between the weak lensing mass and the X-ray
properties.  This paper is a first attempt, but a more comprehensive
study requires a larger sample of clusters. As part of the CCCP we
therefore complement the sample studied here with an additional 30
massive clusters, resulting in a sample of $\sim 50$ clusters with
$T>5$keV, with lensing masses derived from deep CFHT imaging.

The small points in Figure~\ref{lt} correspond to the additional
clusters in the CCCP sample. Although many of these clusters lie on
the same relation as the clusters studied here, the overall scatter is
significantly larger!  Part of the scatter is likely to be caused by
the fact that Horner (2001) did not correct for the presence of cool
cores. For instance, Markevitch (1998) has shown that a considerable
fraction of the scatter in the $L-T$ relation stems from scatter in
the strength of the cool core. This is supported by theoretical models
which indicate that a cool core lowers the average temperature and
raises the X-ray luminosity of a cluster of a given mass (e.g., Voit
et al. 2002; McCarthy et al. 2004). However, the sample of clusters
studied here contains a typical mix of cooling and non-cooling flow
systems similar to other studies (e.g., Peres et al. 1998; Bauer et
al.  2005) and thus we do not expect a bias towards the selection of
clusters with relatively weak cool cores. 

Nevertheless, to understand the wide range of cluster properties,
larger samples of well studied clusters are needed. Such samples will
also improve our estimates of the normalisation and slope of the
mass-temperature and mass-luminosity relations.  The observed slopes
of the mass-temperature relation agree well with theoretical
predictions (e.g., Evrard et al. 1996). However, models that lack
feedback processes do not fare so well for the amplitude. Compared to
these models, X-ray obsevations find lower masses (e.g., Horner et
al. 1999; Nevalainen et al., 2000; Reiprich \& B{\"ohringer, 2002;
Arnaud et al. 2005).  The origin of the differences between theory and
observations are not fully resolved, but several explanations have
been proposed, many of which are related to non-gravitational physics.
Cooling and feedback processes tend to alter the mass-temperature
relation in a systematic way way by raising the gas entropy level (as
compared to the purely gravitational heating scenario). The net result
is a increased temperature, or lower normalisation (e.g., Voit et
al. 2002; Borgani et al. 2004). Because the X-ray mass estimates are
also affected by the complex ICM physics, an independent weak lensing
mass estimate is of great importance if we want to understand the
mass-temperature relation.

Using the expected relation between luminosity and temperature, these
simple models also predict that the mass and luminosity are related
through

\begin{equation}
E(z)M_\Delta=\left(\frac{L_X}{E(z)}\right)^{3/4}.
\end{equation}

Figure~\ref{mass_x}a shows the observed mass-luminosity relation. We
use the value for $M_{2500}$ derived from the aperture mass method.
The best fit power law model is indicated by the dashed
line. Figure~\ref{mass_x}b shows the likelihood contours for the
parameters of this power law model for the $M_{2500}-$luminosity
relation. The best fit model has a $\chi^2=38.1$ for 15 degrees of
freedom. Hence the mass-luminosity shows evidence of intrinsic
scatter. We find a slope of $\alpha=0.43^{+0.09}_{-0.10}$ and a mass
of $M_{2500}=1.94^{+0.11}_{-0.12}\times 10^{14}h^{-1}\msun$ for a
cluster with a luminosity of $10^{45}h^{-2}$ergs/s. The slope expected
from simple self-similar models is ruled out by our measurements at
the $\sim 3\sigma$ level. This is not surprising, as this reflects the
fact that the observed $L-T$ relation is steeper than the one expected
from self-similar models: if we assume $M\propto T^{3/2}$ and $L\propto
T^3$, we expect $M\propto L^{1/2}$, as is observed.

\begin{figure}
\begin{center}
\leavevmode
\hbox{%
\epsfxsize=8.5cm
\epsffile{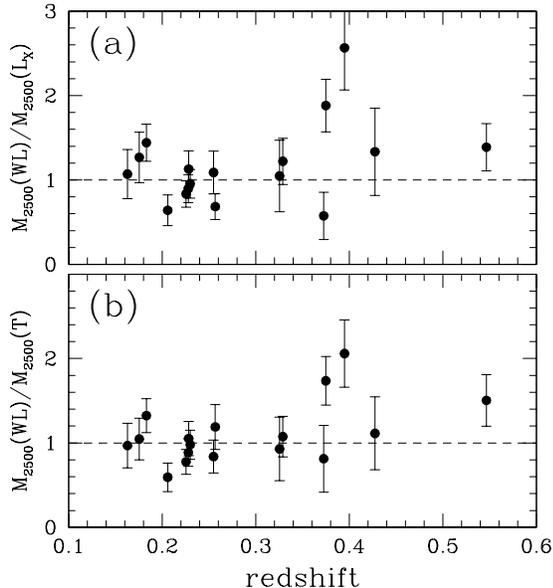}}
\caption{{\it Panel (a):} ratio of $M_{2500}$ from the weak lensing
analysis and the best fit mass-luminosity relation as a function of
redshift. {\it Panel (b):} like a, but now for the best fit
mass-temperature relation. The results suggest that the higher
redshift clusters preferentially have larger than expected lensing
masses, but a larger sample is required to make definite statements.}
\label{evol}
\end{center}
\end{figure}

Figure~\ref{mass_x}c shows the observed mass-temperature relation.  We
find $\chi^2=25.3$ for the best fit power law model, which is somewhat
high, but significantly smaller than for the mass-luminosity relation.
The likelihood contours for the power law model parameters are
presented in Figure~\ref{mass_x}d. In this case we find a slope
$\alpha=1.34^{+0.30}_{-0.28}$, which agrees well with the self-similar
slope of 1.5. The slope also agrees well with studies based on X-ray
data alone (e.g., Nevalainen et al. 2000; Allen et al., 2001;
Arnaud et al. 2005; Vikhlinin et al. 2006).

Our best fit power law model yields a mass of
$M_{2500}=(1.4\pm0.2)\times 10^{14} h^{-1}\msun$ for a cluster with a
temperature of 5~keV. Allen et al. (2001) list a mass of
$M_{2500}=(3.8\pm0.4)\times 10^{14} h^{-1}\msun$ for a 10~keV cluster
(assuming a slope of 1.5). If we fix the slope to 1.5, we find
$M_{2500}=(3.6\pm0.2)\times 10^{14} h^{-1}\msun$ for a 10~keV cluster,
which is in excellent agreement with Allen et al. (2001).

Arnaud et al. (2005) studied the M-T relation of a sample of 6 nearby
relaxed clusters ($T>3.5$~keV), using XMM-Newton. They find a mass of
$(1.3\pm0.04)\times 10^{14}h^{-1}\msun$ for a 5~keV cluster, in
excellent agreement with our findings. Finally, Vikhlinin et
al. (2006) list a mass of $(0.9\pm 0.05)\times 10^{14}h^{-1}\msun$ for
a 5~keV cluster. This result is marginally consistent with our result.
Note, however, that the ASCA temperatures used here may be biased low,
because they have not been corrected for cool cores. A rough
comparison between the Vikhlinin et al. (2006) and Horner (2001)
temperatures suggest that the ASCA temperatures used here are about
10\% lower. This suggests that our normalisation may need to be
reduced by about 15\%.

So far, we have assumed that the evolution of the X-ray properties follows
the expected evolution, parameterized by $E(z)$. To examine whether
the cluster properties show evidence of additional evolution, we plot
the ratio of the lensing mass and the best mass from the best fit
$M-L_X$ relation in Figure~\ref{evol}a. Similarly, Figure~\ref{evol}b
shows the results for the $M-T$ relation. The clusters with $z>0.35$
seem to have somewhat larger than expected lensing masses, but a
larger sample is required to confirm this suggestion. 

We find that the residuals in the $M-L_X$ and $M-T$ relation are
highly correlated. This is not too surprising, given that the clusters
in our sample follow such a tight $L-T$ relation. The implications,
however, are interesting. The largest outliers in Figure~\ref{evol}
are CL0024+16 and A370, for which we find lensing masses that are
$\sim 2$ times larger than expected from the X-ray properties. Yet,
these clusters lies on the $L-T$ relation.  This discrepancy could
indicate a problem with the weak lensing mass determination, but we
were unable to identify an obvious measurement error. However, these
clusters have been studied in detail because of their extreme strong
lensing properties. Therefore, these clusters may be considered
``lensing-selected'', which could explain their larger than
expected masses when compared to their X-ray properties. A related
observation is that both clusters show evidence of recent, or ongoing
merging. The kinematics of CL0024+16 have been studied in detail by
Czoske et al. (2002), who conclude that it might have experienced a
high speed collision. A370 consists of two distinct mass
concentrations (e.g., Kneib et al., 1993), which enhances its strong
lensing cross-section.

\section{Conclusions}

Tremendous progress in measurement techniques, in conjunction with new
field imaging capabilities on large telescopes has lead to the next
step in the study of galaxy clusters using weak gravitational lensing:
the study of the mass distribution of large samples of clusters.  This
paper presents the first results of such a systematic, multiwavelength
study of rich clusters of galaxies.

In this paper we have presented new measurements of the masses of 20
X-ray luminous clusters of galaxies at intermediate redshifts. The
results are based on a careful analysis of deep archival $R$-band data
obtained using the Canada-France-Hawaii-Telescope. In particular, we
accounted for a number of subtle effects that, if ignored, can lead to
small biases or incorrect error estimates. Thanks to the wide field of
view of the CFH12k camera, we were able to derive masses that are
essentially model independent

Comparison of the lensing results with measurements of the velocity
dispersion of cluster galaxies shows good agreement. We typically find
good agreement between our results and weak lensing mass estimates in
the literature. Assuming a power law between the lensing mass and the
X-ray temperature, $M_{2500}\propto T^\alpha$, we find a best fit
slope of $\alpha=1.34^{+0.30}_{-0.28}$. This slope agrees with
self-similar cluster models and studies based on X-ray data alone
(Nevalainen et al. 2000; Allen et al. 2001; Arnaud et al. 2005;
Vikhlinin et al. 2006). For a cluster with a temperature of $kT=5$ keV
we obtain a mass $M_{2500}=(1.4\pm0.2)\times 10^{14}h^{-1}\msun$, in
fair agreement with recent Chandra and XMM studies (e.g., Allen et
al. 2001; Arnaud et al. 2005; Vikhlinin et al. 2006).

The comparison to X-ray properties is complicated by the fact that the
analysis pipelines employed by different groups can yield quite
different luminosities and temperatures. We therefore used
measurements by Horner (2001) which are based on ASCA observations and
not corrected for the presence of cool cores. As part of the Canadian
Cluster Comparison Project, we are re-analysing available modern X-ray
data in a consistent manner. Furthermore, to improve constraints on
the normalisation and slope of the scaling relations between mass and
X-ray properties, as well as quantifying the scatter in these
relations, even larger samples of clusters, with accurate weak lensing
masses, are required. To achieve this goal, we have recently augmented
the sample of clusters studied in this paper with deep $g'$ and $r'$
CFHT Megacam imaging of an additional 30 massive clusters.

\vspace{0.5cm} We thank Andisheh Mahdavi and Arif Babul for many
useful discussions. We are also grateful to Pat Henry and Mark Voit
for comments on the manuscript. This research was supported by the
National Science and Engineering Research Council (NSERC), the
Canadian Foundation for Innovation (CFI), and the Canadian Institute
for Advanced Research (CIAR).  This research used the facilities of
the Canadian Astronomy Data Centre operated by the National Research
Council of Canada with the support of the Canadian Space Agency.

\end{document}